\documentclass[journal,11pt,draftclsnofoot,onecolumn]{IEEEtran}
\usepackage[noadjust]{cite}

\usepackage[T1]{fontenc}
\usepackage{ae}
\usepackage{aecompl}

\ifCLASSINFOpdf
\else
\usepackage[dvips]{graphicx,color}
\fi
\usepackage[cmex10]{amsmath}
\usepackage{multirow}
\usepackage{amsmath,amssymb,bbm,mathrsfs}
\usepackage{type1cm}
\newtheorem{definition}{Definition}[section]
\newtheorem{theorem}{Theorem}[section]
\newtheorem{corollary}{Corollary}[section]
\newtheorem{lemma}{Lemma}[section]

\newtheorem{remark}{Remark}[section]

\newtheorem{example}{Example}[section]
\makeatletter
\renewcommand{\theequation}{\arabic{section}.\arabic{equation}}
\@addtoreset{equation}{section}
\makeatother
\makeatletter
    \renewcommand{\thefigure}{%
    \arabic{section}.\arabic{figure}}
\@addtoreset{figure}{section}
\makeatother
\begin{document}
\title{Second-Order Slepian-Wolf Coding Theorems for Non-Mixed and Mixed Sources}
%
%

\author{Ryo Nomura,~\IEEEmembership{Member,~IEEE,}
        and~Te~Sun~Han,~\IEEEmembership{Life Fellow,~IEEE}
\thanks{R. Nomura is with the School of Network and Information, Senshu University, Kanagawa, Japan,
 e-mail: nomu@isc.senshu-u.ac.jp}
\thanks{T. S. Han is with the National Institute of Information and Communications Technology (NICT), Tokyo, Japan, 
e-mail: han@is.uec.ac.jp }
\thanks{The first author with this work was supported in part by JSPS KAKENHI Grant Number 23760346.}
}

\maketitle

\begin{abstract}
The {\it second-order} achievable rate region in Slepian-Wolf source coding systems is investigated. The  concept of {\it second-order} achievable rates, which enables us to make a finer evaluation of achievable rates, has already been introduced and analyzed for {\it general} sources  in the single-user source coding problem.
Analogously, in this paper, we first define the {\it second-order} achievable rate region for the Slepian-Wolf coding system to establish the source coding theorem in the {\it second-order} sense.
The Slepian-Wolf coding problem for correlated sources is one of typical problems in the multi-terminal information theory. 
In particular, 
Miyake and Kanaya, and Han have established the {\it first-order} source coding theorems for {\it general} correlated sources. 
On the other hand, in general, 
the {\it second-order} achievable rate problem for the Slepian-Wolf coding system with {\it general} sources remains still open up to present.
In this paper we present the analysis concerning the {\it second-order} achievable rates for {\it general} sources which are based on the information spectrum methods developed by Han and Verd\'u. Moreover, we establish the explicit {\it second-order} achievable rate region {\color{black} for i.i.d. correlated sources with {\it countably infinite} alphabets and {\it mixed} correlated sources, respectively, using the relevant asymptotic normality}.
\end{abstract}
%
\IEEEpeerreviewmaketitle
%
%
%
%
\section{Introduction}
%
\IEEEPARstart{W}{e} establish the {\it second-order} source coding theorems for the Slepian-Wolf system.
In the single-user source coding system, Han and Verd\'u \cite{HV93} and Steinberg and Verd\'u \cite{Steinberg} have shown the source coding theorem for {\it general} sources using the {\it information spectrum} methods.  Since the class of {\it general} sources is quite large, their results are very fundamental and useful.
On the other hand, there are several researches concerning a finer evaluation of achievable rates called rates of the {\it second-order}. In variable-length source coding, 
Kontoyiannis \cite{Kon} has established the second-order source coding theorem. 
In channel coding problems, Strassen \cite{Strassen} (see, also Csisz\'{a}r and K\"{o}rner \cite{CK}), Hayashi \cite{Hayashi2}, Polyanskiy, Poor and Verd\'{u} \cite{Polyanskiy2010},
have determined the second-order capacity.
In addition, Hayashi \cite{Hayashi} has given the optimal second-order achievability theorem for the fixed-length source coding problem for {\it general} sources, and actually computed it for i.i.d. sources by invoking the asymptotic normality.
From his analyses we know that the {\it information spectrum} methods \cite{Han} are still effective also for the evaluation on second-order achievable rates.
Nomura and Han \cite{NH2011} have computed the optimal second-order rate for {\it mixed} sources, which is a typical case of {\it nonergodic} sources, again by invoking the relevant asymptotic normality.

In the area of multi-source coding problems, there are various types of source coding problems \cite{Cover}. 
The Slepian-Wolf coding problem for  two correlated sources is one of such typical problems \cite{Cover,SW}. 
In particular, there are two typical settings in the Slepian-Wolf coding problem.
One is the setting that the decoder decodes only the sequence emitted from one source, and the sequence emitted from the other source is used as side-information. The other is the setting that the decoder have to decode both of two sequences emitted from two correlated sources.
We call the former one the full side-information problem.

In the full side-information problem, the {\it first-order} achievable rates for {\it general} correlated sources have been determined by Steinberg and Verd\'u \cite{Steinberg94}. 
On the other hand, Watanabe, Matsumoto and Uyematsu \cite{WMU2010}, and Nomura and Matsushima \cite{Nomura_SITA2009} have considered the {\it second-order} achievable rate problem with full side-information; in this special case, \cite{WMU2010} has given only a \lq\lq necessary" condition on the optimal {second-order} achievable rates for i.i.d. correlated sources, while Nomura and Matsushima \cite{Nomura_SITA2009} with full-side information has exactly determined the {second-order} achievable rates for i.i.d. correlated sources. These results hold also on the basis of information spectrum methods and the asymptotic normality.

In this paper, we consider the ordinary Slepian-Wolf coding problem, {\it not necessarily with} full side-information. In this setting, Miyake and Kanaya \cite{MK} with {\it finite} source alphabets, as well as Han \cite{Han} with {\it countably infinite} source alphabets, has established the  {\it first-order} source coding theorem for {\it general} correlated sources.
The {\it second-order} achievable rate region not necessarily with full side-information for the Slepian-Wolf coding problem was first considered by Nomura and Matsushima \cite{Nomura2011_2}; they have focused only on i.i.d. correlated sources, where, unfortunately, the sufficient condition and the necessary condition do not coincide. 
On the other hand, in this paper, we shall give the necessary and sufficient condition for {\it general} correlated sources with {\it countably infinite} alphabets. Furthermore, for i.i.d. correlated sources with {\it countably infinite alphabets} and mixed i.i.d. correlated sources with {\it finite} alphabets, we apply this fundamental result to derive the explicit {\it second-order} achievable rate region.
 In the {\it second-order} analysis for i.i.d. correlated sources, the multivariate normal distribution function due to the central limit theorem plays the key role, while in the previous literature on the {\it second-order} achievable rates (cf. \cite{WMU2010,Nomura_SITA2009,NH2011}), only one-dimensional normal distribution functions were enough to consider.
Recently, Tan and Kosut \cite{Tan_ISIT2012} has independently established a counterpart of the {\it second-order} source coding theorem for i.i.d. correlated sources, which is derived partly via the method of information spectra in addition to the standard multi-dimensional central limit theorem.
 It should be noted that the result in \cite{Tan_ISIT2012} heavily relies on the method of {\it types}, which necessitates the assumption of {\it finiteness} of source alphabets.

The analyses here are based wholly on the information spectrum methods to invoke the {multi}-dimensional normal distribution functions.
In the {\it second-order} analysis, we extend our results for i.i.d. correlated sources to the {\it mixed} sources consisting of i.i.d. correlated sources.
The class of mixed sources is very important, because all of stationary sources can be regarded as forming mixed sources obtained by mixing stationary ergodic sources with respect to an {\it appropriate probability measure}. The first-order achievable rate region for mixed correlated sources has already been shown by Han \cite{Han} on the basis of the information spectrum methods. We demonstrate in this paper the second-order achievable rate region also by using the information spectrum methods.

The present paper is organized as follows.
In Section \ref{sec:pre}, we define the {\it general} correlated sources and the achievable rate region. Then, we review the previous results on the first-order achievable rate region. In Section \ref{sec:general}, we derive the second-order achievable rate region for {\it general} sources by invoking the {\it information spectrum} methods. In Section \ref{sec:iid}, based on the general formula established in Section \ref{sec:general}, we compute the {second-order} achievable rate region for i.i.d. correlated sources to establish Theorem \ref{main:theo2}, the {\it canonical representation} of which is also given for the first time in this paper. In Section \ref{sec:example}, a comparison of the second-order approach with the error exponent type bound is shown to elucidate the significance of Theorem \ref{main:theo2}. In Section \ref{sec:mixed}, we establish the formula for the second-order achievable rate region for {\it mixed} correlated sources with {\it general mixture}, again by using the relevant asymptotic normality.
Finally, we conclude our results in Section \ref{sec:con}. 
%
%
%
%
%
%
%
\section{Preliminaries} \label{sec:pre}
\subsection{Correlated Sources}
Let ${\cal X}_1$ and ${\cal X}_2$ be alphabets of two correlated sources, where ${\cal X}_1$ and ${\cal X}_2$ may be {\it countably infinite}.
Let $({\bf X}_1,{\bf X}_2) = \{(X_1^n,X_2^n)\}_{n=1}^{\infty}$ denote a general correlated source pair, i.e., $(X_1^n,X_2^n)$ taking values in ${\cal X}_1^n \times {\cal X}_2^n$ is a pair of correlated source variables of block length $n$, and we write as
\[
(X^n_1,X^n_2) = \left( (X_{11},X_{21}),(X_{12},X_{22}),\cdots, (X_{1n},X_{2n}) \right),\]
 and let ${\bf x}_j=x_{j1},x_{j2},\cdots, x_{jn}$ be a realization of random variable $X^n_j \ (j=1,2)$. 
The probability distribution of $({\bf x}_1,{\bf x}_2)$ is denoted by $P_{X_1^nX_2^n}({\bf x}_1,{\bf x}_2)$.
In particular, in the case that the pair of correlated sources has an i.i.d. property, it holds that
\[
P_{X_1^nX_2^n}({\bf x}_1,{\bf x}_2) = \prod_{i=1}^n P_{X_1X_2}(x_{1i},x_{2i}),
\]
with generic correlated random variable $(X_1,X_2)$, where we use the convention that $P_Z(\cdot)$ denotes the probability distribution of $Z$, and $P_{Z|W}(\cdot | \cdot)$ denotes the conditional probability distribution of $Z$ given $W$.
\subsection{$\varepsilon$-Achievable Rate Region}
The fixed-length codes for correlated sources are characterized by a pair of encoders $(\phi_n^{(1)},\phi_n^{(2)})$ and a decoder $\psi_n$.
The encoders are mappings such as $\phi_n^{(1)}:{\cal X}^n_1 \to {\cal M}_n^{(1)}$, $\phi_n^{(2)}:{\cal X}^n_2 \to {\cal M}_n^{(2)}$, where 
\begin{eqnarray*}
{\cal M}_n^{(1)} = \{1,2,\cdots, M_n^{(1)} \}, \ \ \ {\cal M}_n^{(2)} = \{1,2,\cdots, M_n^{(2)} \}
\end{eqnarray*}denote the code sets.
The decoder is defined as a mapping $\psi_n:{\cal M}_n^{(1)} \times {\cal M}_n^{(2)} \to {\cal X}_1^n \times {\cal X}_2^n$. In the sequel, we focus on this Slepian-Wolf type source coding problem.

The performance of fixed-length codes is evaluated in terms of the error probability and the codeword length. 
The error probability is given by
\[
\varepsilon_n = \Pr \{(X^n_1,X^n_2) \neq \psi_n(\phi_n^{(1)}(X_1^n),\phi_n^{(2)}(X_2^n))  \}.
\]
We call such a pair of encoders $(\phi_n^{(1)},\phi_n^{(2)})$ and decoder $\psi_n$ along with error probability $\varepsilon_n$ an $(n, M_n^{(1)},M_n^{(2)}, \varepsilon_n)$ code.

\begin{definition}
A rate pair $(R_1,R_2)$ is called an $\varepsilon$-achievable rate pair if there exists an $(n, M_n^{(1)},M_n^{(2)}, \varepsilon_n)$ code satisfying
\begin{eqnarray*}
\limsup_{n \to \infty} \frac{1}{n} \log M_n^{(1)} \leq R_1 \ \ \mbox{and} \ \  \limsup_{n \to \infty} \frac{1}{n} \log M_n^{(2)} \leq R_2
\end{eqnarray*}
\[
\limsup_{n \to \infty} \varepsilon_n \leq \varepsilon.
\]
\end{definition}
Then, the $\varepsilon$-achievable rate region is defined as the set of all $\varepsilon$-achievable rate pairs:
\begin{definition}($\varepsilon$-Achievable Rate Region)
\begin{eqnarray*}
R(\varepsilon|{\bf X}_1,{\bf X}_2) = \{ (R_1,R_2) | (R_1,R_2)\mbox{ is $\varepsilon$-achievable rate}\}.
\end{eqnarray*}
\end{definition}

For general correlated sources, Miyake and Kanaya \cite{MK} with finite source alphabets have determined the $0$-achievable rate region, while Han \cite{Han} with countably infinite source alphabets have established the $\varepsilon$-achievable rate region for $0 \leq \forall \varepsilon < 1$.
Before describing their results, we need to define the spectral sup-entropy rate as follows\footnote{%
For any sequence $\{Z_n \}_{n=1}^{\infty}$ of real-valued random variables, we define the limit superior in probability of $\{Z_n \}_{n=1}^{\infty}$ by
$\mbox{p-}\limsup_{n \to \infty} Z_n = \inf \left\{ \beta | \lim_{n \to \infty} \Pr \{Z_n > \beta \} = 0 \right\}$ (cf. \cite{Han}) .
}.
\begin{eqnarray*}
\overline{H}({\bf X}_1|{\bf X}_2) & \equiv& \mbox{p-}\limsup_{n \to \infty} \frac{1}{n} \log \frac{1}{P_{X_1^n|X_2^n}(X_1^n|X_2^n)},  \\
\overline{H}({\bf X}_2|{\bf X}_1)& \equiv& \mbox{p-}\limsup_{n \to \infty} \frac{1}{n} \log \frac{1}{P_{X_2^n|X_1^n}(X_2^n|X_1^n)}, \\
\overline{H}({\bf X}_1{\bf X}_2) & \equiv& \mbox{p-}\limsup_{n \to \infty} \frac{1}{n} \log \frac{1}{P_{X_1^nX_2^n}(X_1^nX_2^n)}.
\end{eqnarray*} 

The following {\it first-order} achievable rate theorem reveals the rate region for {\it general} correlated  sources.
\begin{theorem}[Miyake and Kanaya \cite{MK}] \label{theo:1}
 For any {\it general} sources with {\it finite} alphabets, the $0$-achievable rate region $R(0|{\bf X}_1, {\bf X}_2)$ is given as the set of $(R_1,R_2)$ satisfying
\begin{eqnarray*}
R(0|{\bf X}_1,{\bf X}_2) = \{ (R_1,R_2) \left| 
R_1 \geq \overline{H}({\bf X}_1|{\bf X}_2), R_2 \geq \overline{H}({\bf X}_2|{\bf X}_1), \right. R_1+R_2 \geq \overline{H}({\bf X}_1{\bf X}_2).
\}. 
\end{eqnarray*}
\end{theorem}
Han \cite{Han} has generalized their theorem to the case $0 \leq \epsilon <1$ with {\it countably infinite} alphabets;
set the function $F_n(R_1,R_2)$ as
\begin{align*}
{F}_n(R_1,R_2) = & \Pr \left\{ \frac{ 1}{n}\log \frac{1}{P_{X^n_1|X^n_2}(X_1^n|X_2^n)}  \geq R_1 \right. \\
& \hspace*{0.5cm} \mbox { or }  \frac{ 1}{n} \log \frac{1}{P_{X^n_2|X^n_1}(X_2^n|X_1^n)}  \geq R_2  \\
& \hspace*{0.5cm} \mbox{ or } \left. \frac{ 1}{n} \log \frac{1}{P_{X^n_1X^n_2}(X_1^nX_2^n)}  \geq R_1 + R_2 \right\},
\end{align*}
then we have:
\begin{theorem}[Han \cite{Han}] \label{theo:2}
 For any {\it general} sources with countably infinite alphabets, the {\it first-order} $\varepsilon$-achievable rate region is given by
\begin{eqnarray} \label{eq:2-2-1}
R(\varepsilon|{\bf X}_1,{\bf X}_2) = \mbox{Cl} \left(\left\{ (R_1,R_2) \left| \limsup_{n \to \infty} {F}_n(R_1,R_2)  \leq \varepsilon \right.\right\} \right),
\end{eqnarray}
where Cl($\cdot$) denotes the closure operation.
\end{theorem}
This theorem has been established using the following key lemmas:
\begin{lemma}[Han \cite{Han}] \label{lemma1-1}
Let $M_n^{(1)}$ and $M_n^{(2)}$ be arbitrarily given positive integers. Then, for all $n=1,2,\cdots$, there exists an $(n, M_n^{(1)},M_n^{(2)}, \epsilon_n)$ code satisfying
\begin{align*}
\varepsilon_n \leq &  \Pr \left\{ z_n P_{X_1^n|X_2^n}(X_1^n|X_2^n) \leq \frac{1}{M_n^{(1)}} \right. \\
& \hspace*{0.5cm} \mbox{ or } z_n P_{X_2^n|X_1^n}(X_2^n|X_1^n) \leq \frac{1}{M_n^{(2)}}\\
& \hspace*{0.5cm} \left.\mbox{or } z_n P_{X_1^nX_2^n}(X_1^n,X_2^n) \leq \frac{1}{M_n^{(1)}M_n^{(2)}} \right\} + 3z_n,
\end{align*}
where $\{z_n\}_{n=1}^\infty$ is a sequence of arbitrary real numbers such that $z_i >0$ $(\forall i=1,2,\cdots)$. \IEEEQED
\end{lemma}
%
%
\begin{lemma}[Han \cite{Han}] \label{lemma1-2}
Any $(n, M_n^{(1)},M_n^{(2)}, \epsilon_n)$ code satisfies
\begin{align*}
\varepsilon_n  \geq & \Pr \left\{ P_{X_1^n|X_2^n}(X_1^n|X_2^n) \leq \frac{z_n}{M_n^{(1)}}\right. \\
& \hspace*{0.5cm} \mbox{ or } P_{X_2^n|X_1^n}(X_2^n|X_1^n) \leq  \frac{z_n}{M_n^{(2)}} \\
& \hspace*{0.5cm} \left.\mbox{or } P_{X_1^nX_2^n}(X_1^n,X_2^n) \leq \frac{z_n}{M_n^{(1)}M_n^{(2)}} \right\} - 3z_n,
\end{align*}
for all $n=1,2,\cdots,$ where $\{z_n\}_{n=1}^\infty$ is a sequence of arbitrary real numbers such that $z_i >0$ $(\forall i =1,2,\cdots)$. \IEEEQED
\end{lemma}
Notice here that these lemmas are valid for general correlated sources with {\it countably infinite} alphabets.
In this paper too we shall invoke these lemmas in order to derive the {\it second-order} achievable rate region as in the subsequent sections.
%
%
%
%
\subsection{$(a_1,a_2,\varepsilon)$-Achievable Rate Region}
In the single-user source coding problem, a finer evaluation called the {\it second-order} achievable rate 
has been studied. Accordingly, we define the {\it second-order} achievable rate pair as follows.
\begin{definition}
A rate pair $(L_1, L_2)$ is called a {\it second-order} $(a_1, a_2, \varepsilon)$-achievable rate pair if there exists an $(n, M_n^{(1)},M_n^{(2)}, \varepsilon_n)$ code satisfying
\[
\limsup_{n \to \infty} \frac{1}{\sqrt{n}} \log \frac{M_n^{(1)}}{e^{na_1}} \leq L_1 \ \  \mbox{and} \ \ \limsup_{n \to \infty} \frac{1}{\sqrt{n}} \log \frac{M_n^{(2)}}{e^{na_2}} \leq L_2, 
\]
\[
\limsup_{n \to \infty} \varepsilon_n \leq \varepsilon.
\]
\end{definition}
Notice that, in the definition of $(a_1, a_2, \varepsilon)$-achievable rate pairs, the condition for the error probability is as the same as in the definition of $\varepsilon$-achievable rate pair. 
Moreover, let us define the set of $(a_1, a_2, \varepsilon)$-achievable rate pairs given $(a_1,a_2)$ as
\begin{definition}[second-order $(a_1, a_2, \varepsilon)$-achievable rate region]
\begin{eqnarray*}
L( a_1, a_2,\varepsilon|{\bf X}_1,{\bf X}_2) = \{ (L_1,L_2) | (L_1,L_2)\mbox{ is $(a_1, a_2, \varepsilon)$-achievable}\}.
\end{eqnarray*}
\end{definition}
%
%
 %
%
%
%
%
%
%
%
%
\section{$(a_1, a_2, \varepsilon)$-achievable rate region for general correlated sources} \label{sec:general}
We shall determine the $(a_1, a_2, \varepsilon)$-achievable rate region for {\it general} correlated sources with countably infinite alphabets. Before describing our main result, let us define the function $F_n(L_1,L_2|a_1,a_2)$ by 
\begin{align*}
F_n(L_1,L_2|a_1,a_2) = & \Pr \left\{ \frac{-\log P_{X^n_1|X^n_2}(X_1^n|X_2^n) - na_1} {\sqrt{n}} \geq L_1 \right. \\
& \hspace*{0.5cm} \mbox{  or  }\  \frac{-\log P_{X^n_2|X^n_1}(X_2^n|X_1^n) - na_2} {\sqrt{n}} \geq L_2  \\
& \hspace*{0.5cm}  \left. \mbox{ or } \ \frac{-\log P_{X^n_1X^n_2}(X_1^nX_2^n) - n \left(a_1 + a_2\right)} {\sqrt{n}} \geq L_1 + L_2 \right\},
\end{align*}

and set
\begin{align*}
\overline{F}_n(L_1,L_2|a_1,a_2)  \equiv & 1 - F_n(L_1,L_2|a_1,a_2) \\
 = & \Pr \left\{ \frac{-\log P_{X^n_1|X^n_2}(X_1^n|X_2^n) - na_1} {\sqrt{n}} < L_1, \right. \\
 & \hspace*{0.5cm} \frac{-\log P_{X^n_2|X^n_1}(X_2^n|X_1^n) - na_2} {\sqrt{n}} < L_2,  \\
 & \hspace*{0.5cm}  \left. \frac{-\log P_{X^n_1X^n_2}(X_1^nX_2^n) - n \left(a_1 + a_2\right)} {\sqrt{n}} < L_1 + L_2 \right\}.
\end{align*}
It should be remarked that the function $\overline{F}_n(L_1,L_2|a_1,a_2)$ thus defined  is a multivariate cumulative distribution function.
Now, we have
\begin{theorem} \label{theo:main1}
\begin{align} \label{fn}
L(a_1, a_2,\varepsilon|{\bf X}_1,{\bf X}_2) = \mbox{Cl} \Bigl( \Bigl\{ (L_1,L_2) & \left| \limsup_{n \to \infty} F_n(L_1,L_2|a_1,a_2) \leq  \varepsilon \right. \Bigr\} \Bigr).  
\end{align}
\end{theorem}
\begin{IEEEproof}See Appendix A.
\end{IEEEproof}
This theorem provides us with the information spectrum basis for establishing Theorem \ref{main:theo2} (for i.i.d. sources) and Theorems \ref{theo:mix1}, \ref{theo:mix2} (for mixed sources) later.
%
%
%
%
%
%
%
\section{$(a_1, a_2, \varepsilon)$-achievable rate region for i.i.d. correlated sources} \label{sec:iid}
Although Theorem \ref{theo:main1} is valid for {\it general} correlated sources, it is hard to actually compute it. 
In the subsequent sections, we compute the $(a_1, a_2, \varepsilon)$-achievable rate region for several typical cases of theoretical/practical importance.
To do so, first in this section we focus on i.i.d. correlated sources. 

\begin{figure}[h]
\begin{center}
\includegraphics[width=2.6in]{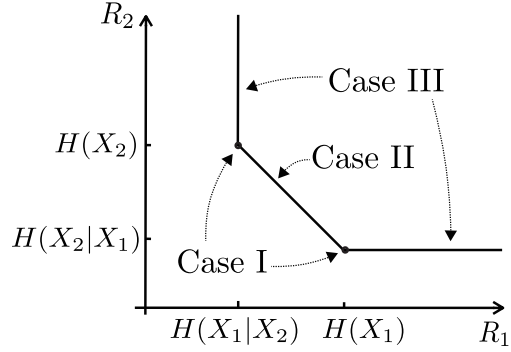}
\centering \caption{Boundary Points.}
\label{fig_boundary}
\end{center}
\end{figure}

In this case, Theorem \ref{theo:1}, as is well known, reduces to the first-order Slepian-Wolf theorem for i.i.d. correlated sources, i.e., the Slepian-Wolf region:
\begin{eqnarray*}
R(0|{\bf X}_1,{\bf X}_2) = \{ (R_1,R_2) \left| 
R_1 \geq {H}({X}_1|{X}_2), R_2 \geq {H}({X}_2|{X}_1), \right. R_1+R_2 \geq {H}({X}_1{X}_2)
\},
\end{eqnarray*}
with generic correlated source variable $(X_1,X_2)$, the conditional entropies $H(X_1|X_2)$, $H(X_2|X_1)$ and the joint entropy $H(X_1X_2)$, which forms the polygon with the following boundary points (see Fig. \ref{fig_boundary}): 
\vspace*{.5\baselineskip}
\\
{Case I (Corner Points)}: $$a_1 = H(X_1|X_2) \ \ \mbox{ and } \ \ a_2 = H(X_2);$$
$$a_1 = H(X_1) \ \ \mbox{ and } \ \ a_2 = H(X_2|X_1),$$ \\
Case II (Non-Corner Points): For correlated $X_1$, $X_2$ and $0 < \forall \lambda <1 $,
$$a_1 = \lambda H(X_1) + (1-\lambda) H(X_1|X_2)\ \ \mbox{ and } \ \ a_2 = (1-\lambda) H(X_2) + \lambda H(X_2|X_1),$$ \\
Case III (Full Side Points):
$$a_1 = H(X_1|X_2)\ \ \mbox{ and } \ \ a_2 > H(X_2);$$
$$a_1 > H(X_1)\ \ \mbox{ and } \ \ a_2 = H(X_2|X_1).$$

In each of these cases, we compute the $(a_1, a_2, \varepsilon)$-achievable rate region on the basis of Theorem \ref{theo:main1}.
\begin{remark} \label{remark:4-1}
It is not difficult to check that $(L_1,L_2)$ can take arbitrary values in $\mathbb{R}^2$ ($\mathbb{R} =$ the set of all real numbers ) if $(a_1,a_2)$ is an internal point of the polygon, while the set of all achievable rate pairs $(L_1, L_2)$ reduces to the empty set $\emptyset$ if $(a_1,a_2)$ is outside the polygon. Thus, we may focus only on the above cases. \IEEEQED
\end{remark}

Now, we define 
\begin{align*}
\Phi(T_1,T_2,T_3) \equiv &  \lim_{n \to \infty} \Pr \left\{ \frac{-\log P_{X^n_1|X^n_2}(X_1^n|X_2^n) - nH(X_1|X_2)} {\sqrt{n}} < T_1, \right. \\
& \ \ \ \ \  \frac{-\log P_{X^n_2|X^n_1}(X_2^n|X_1^n) - nH(X_2|X_1)} {\sqrt{n}} < T_2,   \\
& \ \ \ \ \  \left. \frac{-\log P_{X^n_1X^n_2}(X_1^nX_2^n) - n H(X_1X_2)} {\sqrt{n}} < T_3 \right\},
\end{align*}
then, by means of the multi-dimensional central limit theorem based on the i.i.d. property of sources (cf. Feller \cite{Feller}, Sazonov \cite{Sazonov}, Bentkus \cite{Bentkus}), we see that $\Phi(T_1,T_2,T_3)$ specifies a three-dimensional normal cumulative distribution function; more specifically,
\[
\Phi(T_1,T_2,T_3) \equiv  \int^{T_1}_{-\infty} dy_1 \int^{T_2}_{-\infty} dy_2 \int^{T_3}_{-\infty} d y_3 \frac{1}{(\sqrt{2\pi})^3 \sqrt{\det \Sigma}} \exp \left( -\frac{1}{2} {\bf y}  \Sigma^{-1} {\bf y}^{\mathrm{T}}  \right),
\]
where ${\bf y}=(y_1,y_2,y_3)$ is a three-dimensional row vector, and $\Sigma= (\sigma^2_{ij})\ \ (i,j = 1,2,3)$ denotes the dispersion matrix  (cf. Tan and Kosut \cite{Tan_ISIT2012}) i.e., the covariance matrix given by
\[
\sigma^2_{ij} = \sum_{x_1 \in {\cal X}_1}\sum_{x_2 \in {\cal X}_2} P_{X_1X_2}(x_1,x_2) z_i(x_1,x_2)z_j(x_1,x_2),
\]
where
\[
z_{1}(x_1,x_2) = \log \frac{1}{P_{X_1|X_2}(x_1|x_2)} - H(X_1|X_2),
\]
\[
z_{2}(x_1,x_2) = \log \frac{1}{P_{X_2|X_1}(x_2|x_1)} - H(X_2|X_1),
\]
\[
z_{3}(x_1,x_2) = \log \frac{1}{P_{X_1X_2}(x_1,x_2)} - H(X_1X_2).
\]
In order to avoid subtle irregularities, we assume throughout that $\Sigma$ is positive-definite (As for a treatment of the case with singular $\Sigma$, see \cite{Tan_ISIT2012}).

We then consider the marginal cumulative distributions $\Phi_{23}(T_2,T_3)$, $\Phi_{3}(T_3)$ of $\Phi(T_1,T_2,T_3)$ such that 
\[
\Phi_{23}(T_2,T_3) \equiv \lim_{T_1 \to \infty }\Phi(T_1, T_2,T_3),
\]
\[
\Phi_{3}(T_3) \equiv \lim_{T_1,T_2 \to \infty} \Phi(T_1, T_2,T_3).
\]
Other marginal cumulative distributions can be defined in similar manners.
Notice that $\Phi_{3}(T_3)$ and (resp. $\Phi_{23}(T_2,T_3)$) are one-dimensional (resp. two-dimensional) normal cumulative distribution functions.

\begin{figure}[t]
\begin{center}
\includegraphics[width=2.6in]{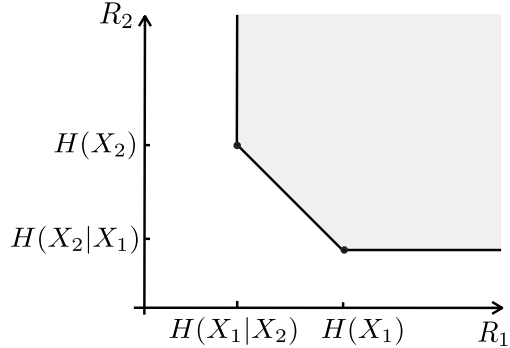}
\centering \caption{First-Order Achievable Rate Region ($\varepsilon =0$)}
\label{fig_1}
\end{center}
\end{figure}

\begin{figure}[t]
\begin{center}
\includegraphics[width=3in]{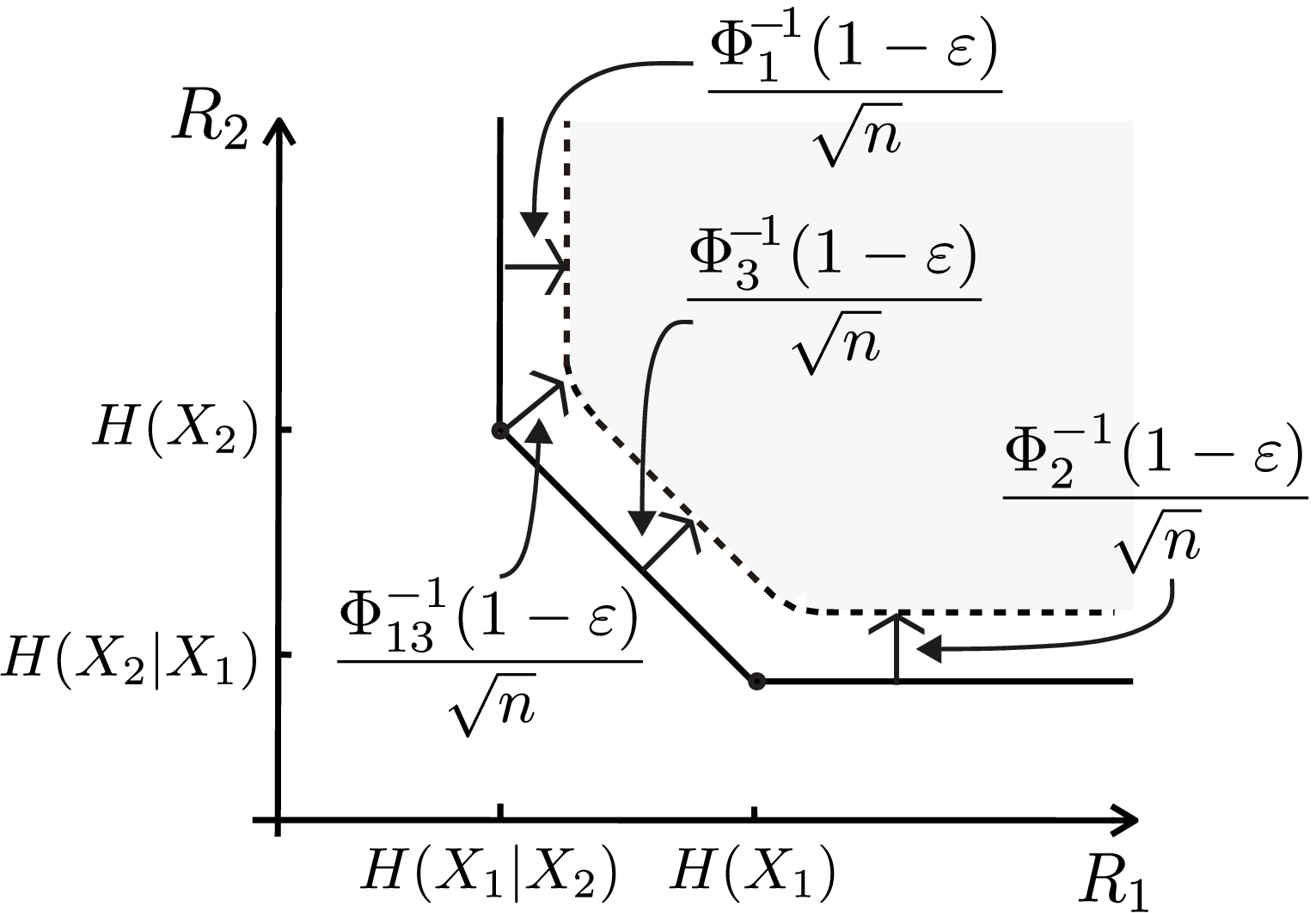}
\centering \caption{Second-Order Achievable Rate Region for Case I, Case II and Case III ($\varepsilon < 1/2$)}
\label{fig_2}
\end{center}
\end{figure}

\begin{figure}[t]
\begin{center}
\includegraphics[width=3in]{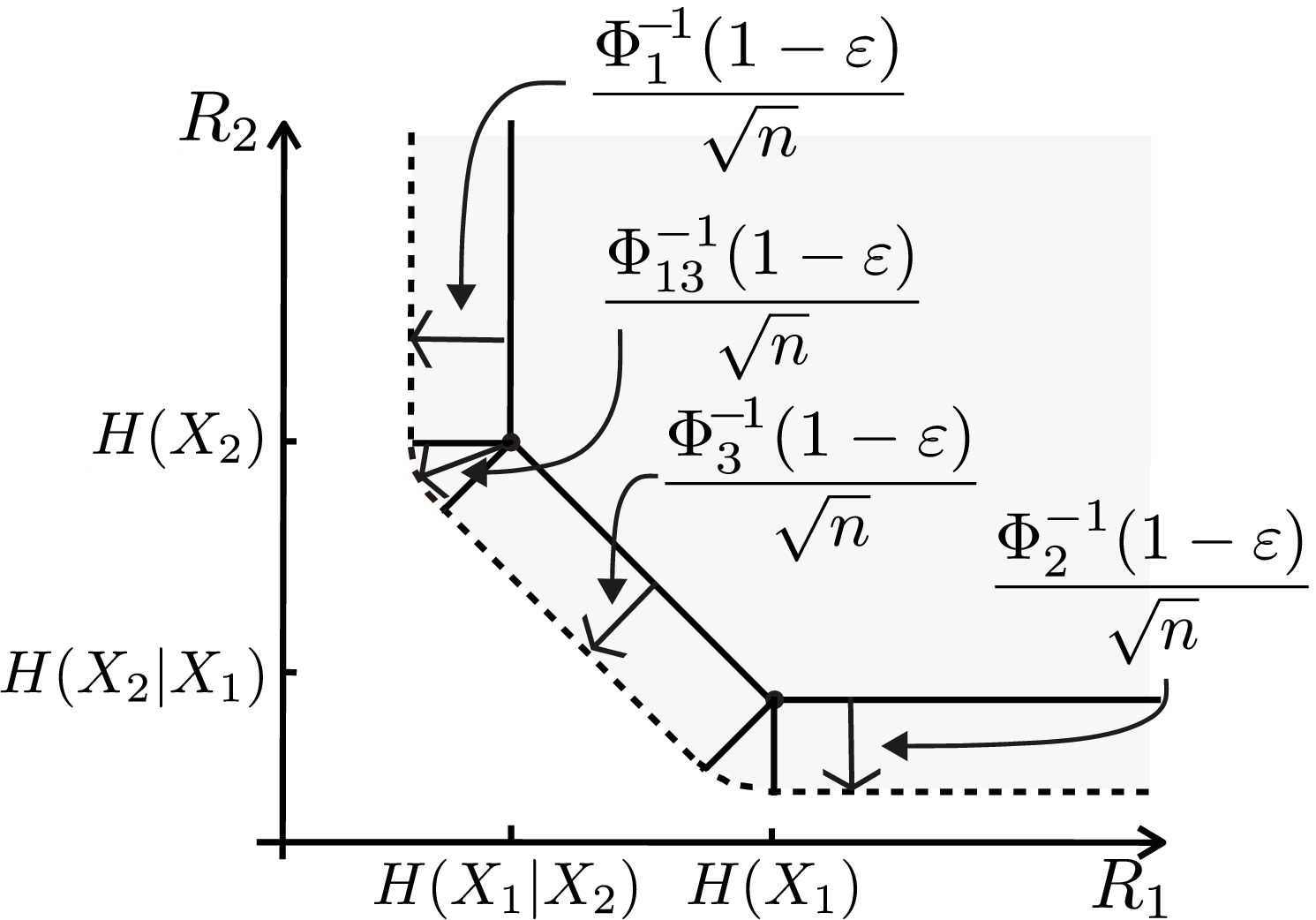}
\centering \caption{Second-Order Achievable Rate Region for Case I, Case II and Case III ($\varepsilon > 1/2$)}
\label{fig_3}
\end{center}
\end{figure}

Now, we have the following theorem: 
\begin{theorem} \label{main:theo2}
For any i.i.d. correlated sources with {\it countably infinite} alphabets, we have for all $0 \leq \varepsilon < 1$: 
\vspace*{\baselineskip}
\\
{\it Case I:}  $a_1 =H(X_1|X_2)$ and $a_2 = H(X_2)$ (without loss of generality): 
\begin{eqnarray*} 
L(a_1, a_2, \varepsilon|{\bf X}_1,{\bf X}_2) = \left\{ (L_1, L_2) \left| \Phi_{13} \left( L_1, L_1+L_2 \right) \geq 1 - \varepsilon  \right.  \right\}.
\end{eqnarray*}
{\it Case II:} for correlated $X_1$, $X_2$ and  $0 < \forall \lambda < 1$: 
\begin{eqnarray*} 
L(a_1, a_2, \varepsilon|{\bf X}_1,{\bf X}_2) =  \left\{ \left(L_1, L_2\right) \left| \Phi_{3} \left( L_1+L_2 \right) \geq 1- \varepsilon \right. \right\}.
\end{eqnarray*} 
{\it Case III:} $a_1 =H(X_1|X_2)$ and $a_2 > H(X_2)$ (without loss of generality): 
\begin{eqnarray*} 
L(a_1, a_2, \varepsilon|{\bf X}_1,{\bf X}_2) =  \left\{ (L_1, L_2) \left| \Phi_{1} \left( L_1 \right) \geq 1- \varepsilon \right. \right\}.
\end{eqnarray*} 
\end{theorem}
\begin{IEEEproof}
See Appendix B.
\end{IEEEproof}
%
%
%
%
\begin{remark} \label{remark:4-M}
Here, in view of the definitions of first-order and second-order achievable rates, any achievable rates can be expressed as in the form of (first-order achievable rate) $+$ $\frac{1}{\sqrt{n}}$(second-order achievable rate). Thus, the two-dimensional set ${\cal A}_n(\varepsilon)$ of all achievable rate pairs 
$
\left( 
R_1^{(n)},
R_2^{(n)}
\right)^{\mathrm{T}}
$
in the finite blocklength regime turns out to be expressed, up to the second-order, as 
\begin{equation} \label{eq:4-M}
{\cal A}_n(\varepsilon) = \bigcup_{(a_1,a_2)^{\mathrm{T}} \in R(0|{\bf X}_1,{\bf X}_2)} \left\{ \left( \begin{array}{c}
a_1 \\
a_2
\end{array}
\right)
+ \frac{1}{\sqrt{n}} L(a_1, a_2, \varepsilon | {\bf X}_1, {\bf X}_2)  \right\},
\end{equation}
where the first-order Slepian-Wolf region with $\varepsilon =0$ is depicted in Fig. \ref{fig_1}. Then, with the aid of Theorem \ref{main:theo2}, the boundary $\partial {\cal A}_n(\varepsilon)$ of ${\cal A}_n(\varepsilon)$ is also depicted in Fig. \ref{fig_2} and Fig. \ref{fig_3} with broken lines, i.e., the boundaries $\partial {\cal A}_n(\varepsilon)$ for the cases of $\varepsilon <\frac{1}{2}$ and $\varepsilon > \frac{1}{2}$, respectively, are depicted there. It should be noted that if $\varepsilon > \frac{1}{2}$ then the second-order achievable rates $L_1, L_2$ can be negative, i.e., in all of Cases I, II and III in Theorem \ref{main:theo2} with $\varepsilon > \frac{1}{2}$ the optimal achievable rates 
$
\left( 
R_1^{(n)},
R_2^{(n)}
\right)^{\mathrm{T}}
$
in the finite blocklength regime approach (when $n$ becomes large) the optimal first-order achievable rate region $R(0|{\bf X}_1, {\bf X}_2)$ {\it from outside}.
\end{remark}
%
%
%
%
%
%
%
\begin{remark}
Tan and Kosut \cite{Tan_ISIT2012} defined with {\it finite} source alphabets the region ${\cal R}_{\mathrm{in}}(n, \varepsilon) \subset \mathbb{R}^2$ to be the set of rate pairs $(R_1, R_2)$ such that
\begin{IEEEeqnarray}{rCl} \label{eq:Tan1}
{\bf R} \geq {\bf H} + \frac{1}{\sqrt{n}}{\mathscr S}(\Sigma, \varepsilon),
\end{IEEEeqnarray}
where \lq\lq $\geq$" means componentwise inequality, and 
\[
{\bf R} = (R_1, R_2, R_1+R_2)^{\mathrm{T}}, \quad {\bf H} = (H(X_1|X_2), H(X_2|X_1), H(X_1X_2))^{\mathrm{T}},
\]
\begin{IEEEeqnarray*}{rCl}
{\mathscr S}(\Sigma, \varepsilon) = \left\{ {\bf z} \in \mathbb{R}^3 \left| \Pr \left\{ {\bf Z} \leq {\bf z}  \right\} \geq 1 - \varepsilon \right. \right\}
\end{IEEEeqnarray*}
with ${\bf Z} \sim {\cal{N}}({\bf 0},\Sigma)$; then they proved that ${\cal R}_{\mathrm{in}}(n, \varepsilon)$ is a set of achievable rate pairs up to the second-order. It is not difficult to check that ${\cal R}_{\mathrm{in}}(n, \varepsilon) \!=\! {\cal A}_n(\varepsilon)$ for {\it finite} source alphabets. It should be remarked here that inequality  \lq\lq $\geq$" in (\ref{eq:Tan1}) was  \lq\lq $\in$" in the original formula of \cite{Tan_ISIT2012}. However, the use of  \lq\lq $\in$" is misleading, whereas  \lq\lq $\geq$" gives the right  notation. \IEEEQED
\end{remark}
\begin{remark} \label{remark:4-2}
Notice that both of $\Phi_{3} \left( L_1+L_2 \right) $ and $\Phi_{1} \left( L_1\right)$ are one-dimensional normal cumulative distribution functions.
Thus, they can be simply written as 
\[
\Phi_{3} \left( L_1+L_2 \right) = \int^{L_1+L_2}_{-\infty} \frac{1}{\sqrt{2\pi} \sigma_{33}}  \exp \left( -\frac{y_3^2}{2\sigma^2_{33}} \right) d{y_3},
\]
\[
\Phi_{1} \left( L_1\right) = \int^{L_1}_{-\infty} \frac{1}{\sqrt{2\pi} \sigma_{11}}  \exp \left( -\frac{y_1^2}{2\sigma^2_{11}} \right) d{y_1},
\]
and, similarly
\[
\Phi_{13} \left( L_1,L_1+L_2 \right) = \int^{L_1}_{-\infty} dy_1 \int^{L_1+L_2}_{-\infty} dy_3 \frac{1}{{2\pi} \sqrt{\det \Sigma_{13}}}  \exp \left( -\frac{1}{2}{\bf y}_{13}\Sigma_{13}^{-1} {\bf y}_{13}^{\mathrm{T}} \right),
\]
where
\[
\Sigma_{13} = \left( \begin{array}{cc}
\sigma^2_{11} & \sigma^2_{13} \\
\sigma^2_{31} & \sigma^2_{33} \\
\end{array} \right), \ \ 
{\bf y}_{13} = (y_1,y_3).
\]
Thus, the {\it second-order} achievable rate region in Case II reduces simply to $L_1+L_2 \geq T_{\rm II}$ where $T_{\rm II}$ is specified by 
\[
\int^{T_{\rm II}}_{-\infty} \frac{1}{\sqrt{2\pi} \sigma_{33}}  \exp \left( -\frac{y_3^2}{2\sigma^2_{33}} \right) d{y_3} = 1 -\varepsilon \quad \left( T_{\rm II} = \Phi_{3}^{-1}(1-\varepsilon) \right);
\]
and the {\it second-order} achievable rate region in Case III reduces simply to $L_1 \geq T_{\rm III}$ ($L_2$ is arbitrary) where $T_{\rm III}$ is specified by
\[
\int^{T_{\rm III}}_{-\infty} \frac{1}{\sqrt{2\pi} \sigma_{11}}  \exp \left( -\frac{y_1^2}{2\sigma^2_{11}} \right) d{y_1} = 1 -\varepsilon \quad \left( T_{\rm III} = \Phi_{1}^{-1}(1-\varepsilon) \right).
\]
On the other hand, the {\it second-order} achievable rate region in Case I is explicitly written as the set of all $(L_1,L_2)$ such that 
\[
\Phi_{13} \left( L_1,L_1+L_2 \right) = \int^{L_1}_{-\infty} dy_1 \int^{L_1+L_2}_{-\infty} dy_3 \frac{1}{{2\pi} \sqrt{\det \Sigma_{13}}}  \exp \left( -\frac{1}{2}{\bf y}_{13}\Sigma_{13}^{-1} {\bf y}_{13}^\mathrm{T} \right) \geq 1 - \varepsilon.
\]
\end{remark}
\begin{remark}
Notice that Case III is substantially the same as the full side-information problem. 
Actually, in Case III, there is no condition on $L_2$ for the achievable rate region. This means that the error probability is independent of the second-order rate $L_2$. This is because, for the decoder, the first-order rate $a_2 > H(X_2)$ is sufficient to reconstruct ${X_2^n = {\bf x}_2}$.
Therefore, the decoder is able to decode ${X_2^n = {\bf x}_2}$ perfectly without the knowledge of $X_1^n = {\bf x_1}$. Then, using the ${X_2^n = {\bf x}_2}$ the decoder decodes $X_1^n = {\bf x_1}$. 
That is, ${X_2^n = {\bf x}_2}$ can be regarded as providing full side-information.
Consequently, the result of Case III coincides with the results in \cite{WMU2010,Nomura_SITA2009}, in which  full side-information problems are treated. \QED
\end{remark}

Theorem \ref{main:theo2} can be equivalently restated as the following corollary: let
\begin{align} \label{eq:solution1}
K_n(a_1, a_2,\varepsilon|{\bf X}_1,{\bf X}_2)  \equiv & \Bigl\{ (L_1,L_2)  \left| \Phi \left(\sqrt{n}\left(a_1- H(X_1|X_2)\right) + L_1, \right. \right. \nonumber \\
& \hspace*{0.5cm} \sqrt{n}\left(a_2 - H(X_2|X_1) \right) + L_2, \nonumber \\
& \hspace*{0.5cm}  \left.\sqrt{n}\left(a_1+a_2 - H(X_1X_2) \right) + L_1+L_2 \right) \geq  1 - \varepsilon \Bigr\},
\end{align}
then, we have:
\begin{corollary} \label{coro1}
For an i.i.d. correlated sources with {\it countably infinite} alphabets,
the second-order $(a_1, a_2, \varepsilon)$-achievable rate region is given as the two-dimensional set\footnote{%
Generally speaking, $\lim_{n \to \infty}A_n$ denotes the {\it limit} of a sequence $\{A_n \}_{n=1}^\infty$ of sets $A_n$'s in the sense that $\lim_{n \to \infty} A_n := \limsup_{n \to \infty} A_n = \liminf_{n \to \infty} A_n$ (cf. Billingsley \cite{Billingsley})}
:
\begin{IEEEeqnarray}{rCl} \label{eq:coro1}
L(a_1, a_2,\varepsilon|{\bf X}_1,{\bf X}_2) = \lim_{n \to \infty} K_n(a_1, a_2,\varepsilon|{\bf X}_1,{\bf X}_2).
\end{IEEEeqnarray}
\end{corollary}
\begin{remark}
It is easy to see that the right-hand side of (\ref{eq:coro1}) is rewritten as 
\begin{IEEEeqnarray}{rCl} \label{canonical}
\lefteqn{\lim_{n \to \infty} K_n(a_1, a_2,\varepsilon|{\bf X}_1,{\bf X}_2)} \nonumber \\
& = & \left\{ (L_1,L_2 \right) \left| \lim_{n \to \infty} \Phi(\sqrt{n}(a_1 - H(X_1|X_2)) + L_1, \right.  \nonumber \\
&& \hspace*{1.8cm} \sqrt{n}(a_2-H(X_2|X_1)) + L_2, \nonumber \\
& &\hspace*{1.8cm} \left. \sqrt{n}(a_1+a_2 - H(X_1X_2)) + L_1+L_2) \geq 1-\varepsilon \right\}.
\end{IEEEeqnarray} 
\ \QED
\end{remark}

\begin{IEEEproof}[Proof of Corollary \ref{coro1}]
It is not difficult to verify by letting $n \to \infty$ that $K_n(a_1, a_2,\varepsilon|{\bf X}_1,{\bf X}_2)$ asymptotically coincides with each case in Theorem \ref{main:theo2} with $a_1$ and $a_2$ as in Cases I, II and III, respectively. Moreover, Corollary \ref{coro1} also includes trivial cases such that $(a_1, a_2)$ is inside or outside the polygon (cf. Remark \ref{remark:4-1}). Thus, it is concluded that Theorem \ref{main:theo2} along with Remark \ref{remark:4-1} is equivalent to Corollary \ref{coro1}.
\end{IEEEproof}
\begin{remark}
Compared with Theorem \ref{main:theo2}, it is observed that the description of Corollary \ref{coro1} needs only a {\it single} equation (\ref{eq:coro1}) but no longer any classifications of pairs $(a_1,a_2)$.
This is a great advantage of the information spectrum methods. 
Thus, we call (\ref{eq:coro1}) the {\it canonical representation} for $L(a_1, a_2,\varepsilon|{\bf X}_1,{\bf X}_2)$.
This point of view is inherited also to Section \ref{sec:mixed}, which enables us to successfully establish the second-order achievable rate region for {\it mixtures} of correlated i.i.d. sources.
This approach is completely different from that of Tan and Kosut \cite{Tan_ISIT2012}.
\end{remark}
\begin{remark}
Let us consider with (\ref{eq:solution1}) the following equation for $L_1$, $L_2$, given $a_1$, $a_2$, $\varepsilon$, $n$:
\begin{IEEEeqnarray}{rCl} \label{eq:4-K}
\Phi &&  \left(\sqrt{n}\left(a_1- H(X_1|X_2)\right) + L_1, \right. \nonumber \\
 && \hspace*{0.3cm} \sqrt{n}\left(a_2 - H(X_2|X_1) \right) + L_2, \nonumber \\
 && \hspace*{0.3cm} \left.\sqrt{n}\left(a_1+a_2 - H(X_1X_2) \right) + L_1+L_2 \right)  = 1 - \varepsilon,
\end{IEEEeqnarray}
and denote the solution (as $n$ tends to $\infty$) by $L_1^{\ast}(a_1,a_2,\varepsilon)$, $L_2^{\ast}(a_1,a_2,\varepsilon)$, respectively. Then, it is not difficult to verify that 
\begin{equation*}
\partial {\cal A}_n(\varepsilon) = \bigcup_{(a_1,a_2)^{\mathrm{T}} \in \partial_{\mathrm{SW}}} \left\{ \left( \begin{array}{c}
a_1 \\
a_2
\end{array}
\right)
+ \frac{1}{\sqrt{n}} 
\left( \begin{array}{c}
L_1^{\ast}(a_1,a_2,\varepsilon) \\
L_2^{\ast}(a_1,a_2,\varepsilon)
\end{array} \right)
 \right\},
\end{equation*}
where $\partial_{\mathrm{SW}}$ is the boundary of the Slepian-Wolf region $R(0|{\bf X}_1, {\bf X}_2)$, and $\partial {\cal A}_n(\varepsilon)$ is the boundary of ${\cal A}_n(\varepsilon)$ defined as in Remark \ref{remark:4-M}. \IEEEQED
\end{remark}
\begin{remark}
It is interesting to note that in Cases II and III the condition on $(a_1, a_2, \varepsilon)$-achievable rate region is described by one-dimensional normal distribution function, while in Case I that is described by two-dimensional normal distribution (cf. Fig. \ref{fig_2} and Fig. \ref{fig_3}).
\end{remark}
%
%
%
%
%
%
%
%
\section{Comparison with the Koshelev Bound and Numerical Examples }   \label{sec:example}
In this section, we give some examples of the second-order achievable rate region for i.i.d. binary correlated sources to elucidate the effectiveness of Theorem \ref{main:theo2}.
We compare the region so far derived with the {\it modified} Koshelev bound (\cite{Koshelev}), i.e., the Gallager type bound for the Slepian-Wolf source coding system.
In Hayashi \cite[Sect. V]{Hayashi2}, the optimal second-order capacity rate in channel coding has been compared with the Gallager bound \cite{Gallager} from the viewpoint of error probabilities vs. achievable rates.
Analogously, in this section we compare the error probability guaranteed by Theorem \ref{main:theo2} with the Koshelev type of error exponent that is  derived on the basis of the maximum likelihood rule.

Let 
\begin{IEEEeqnarray*}{rCl}
E_{1}(s_1) & \equiv & - \log \sum_{{ x}_2 \in {\cal X}_2} \left(  \sum_{{ x}_1 \in {\cal X}_1} P_{X_1X_2}({ x}_1, { x}_2)^{\frac{1}{1+s_1}} \right)^{1+s_1} \\
E_{2}(s_2) &\equiv & - \log \sum_{{ x}_1 \in {\cal X}_1} \left(  \sum_{{ x}_2 \in {\cal X}_2} P_{X_1X_2}({x}_1, { x}_2)^{\frac{1}{1+s_2}} \right)^{1+s_2} \\
E_{3}(s_3) &\equiv & - \log \left( \sum_{({ x}_1,{ x}_2) \in {\cal X}_1 \times {\cal X}_2} P_{X_1X_2}({ x}_1, { x}_2)^{\frac{1}{1+s_3}} \right)^{1+s_3},
\end{IEEEeqnarray*}
where it is evident that $E_1(0) = E_2(0) = E_3(0) = 0$.
Then, we have the following lemma, which is a stronger version of the original Koshelev bound \cite{Koshelev} (Notice that, on the contrary to here, $s_1, s_2, s_3$ are constrained so as to be $0 \leq s_1 = s_2 = s_3 \leq 1$ in \cite{Koshelev}):
\begin{lemma} \label{exponent}
Let $R_1 = \frac{1}{n}\log {M_n^{(1)}}$ and $R_2 = \frac{1}{n}\log{M_n^{(2)}}$, then there exists an $(n, M_n^{(1)},M_n^{(2)}, \varepsilon_n)$ code satisfying
\begin{IEEEeqnarray*}{rCl}
\varepsilon_n & \leq & \min_{0 \leq s_1 \leq 1} \exp\left[ -n\left(R_1s_1 \!-\! E_{1}(s_1)\right) \right] + \min_{0 \leq s_2 \leq 1} \exp\left[ -n\left( R_2s_2 - E_{2}(s_2)\right) \right]  \\
&&+ \min_{0 \leq s_3 \leq 1} \exp\left[ -n \left( \left(R_1+R_2 \right)s_3  - E_{3}(s_3) \right) \right].
\end{IEEEeqnarray*}
\end{lemma}
\begin{IEEEproof}
See Appendix C.
\end{IEEEproof}
We call this bound merely the Koshelev bound for simplicity.
Achievable rate pairs $(R_1, R_2)$ close to the boundary $\partial_{\mathrm{SW}}$ of the Slepian-Wolf region are classified into the following three cases:
\vspace*{.5\baselineskip}
\\
{Case I (Corner Points)}: $$R_1 = H(X_1|X_2) + \frac{L_1}{\sqrt{n}} \ \ \mbox{ and } \ \ R_2 = H(X_2) + \frac{L_2}{\sqrt{n}},$$ \\
Case II (Non-Corner Points): For correlated $X_1$, $X_2$ and $0 < \forall \lambda <1 $,
$$R_1 = \lambda H(X_1) + (1-\lambda) H(X_1|X_2) + \frac{L_1}{\sqrt{n}}\ \ \mbox{ and } \ \ R_2 = (1-\lambda) H(X_2) + \lambda H(X_2|X_1) + \frac{L_2}{\sqrt{n}},$$ \\
Case III (Full Side Points):
$$R_1 = H(X_1|X_2)+ \frac{L_1}{\sqrt{n}} \ \ \mbox{ and } \ \ R_2 > H(X_2)+\frac{L_2}{\sqrt{n}}.$$

In each of these cases, we calculate the Koshelev bound for comparison. These calculations are the Slepian-Wolf source coding counterpart of that of Hayashi \cite[Sect. V]{Hayashi2} for channel coding.

In Case I: the Koshelev bound is given by
\begin{IEEEeqnarray}{rCl} \label{gallager1}
\varepsilon_n & \leq & \min_{0 \leq s_1 \leq 1} \exp\left[ - n\left(\left(H(X_1|X_2) + \frac{L_1}{\sqrt{n}}\right)s_1 \!-\! E_{1}(s_1)\right) \right]  \nonumber \\
&& + \min_{0 \leq s_2 \leq 1} \exp\left[ - n\left( \left(H(X_2) + \frac{L_2}{\sqrt{n}}\right)s_2 - E_{2}(s_2)\right) \right] \nonumber \\
&&+ \min_{0 \leq s_3 \leq 1} \exp\left[ - n \left( \left(H(X_1X_2) + \frac{L_1+L_2}{\sqrt{n}}\right)s_3  - E_{3}(s_3) \right) \right].
\end{IEEEeqnarray} 
In view of 
\begin{equation}\label{eq:exponent2}
\left. \frac{d E_{1}(s_1)}{d s_1} \right|_{s_1=0} = H(X_1|X_2)
,\ \ \left. \frac{d^2 E_{1}(s_1)}{d^2 s_1} \right|_{s_1=0} = \sigma_{11}^2
\end{equation}
\begin{equation}\label{eq:exponent3}
\left. \frac{d E_{2}(s_2)}{d s_2} \right|_{s_2=0} = H(X_2|X_1)
,\ \ \left. \frac{d^2 E_{2}(s_2)}{d^2 s_2} \right|_{s_2=0} = \sigma_{22}^2
\end{equation}
\begin{equation}\label{eq:exponent4}
\left. \frac{d E_{3}(s_3)}{d s_3} \right|_{s_3=0} = H(X_1X_2)
,\ \ \left. \frac{d^2 E_{3}(s_3)}{d^2 s_3} \right|_{s_3=0} = \sigma_{33}^2,
\end{equation}
the first and the third term on the right-hand side of (\ref{gallager1}) are given by
\begin{IEEEeqnarray}{rCl} \label{eq:exponent}
\min_{0 \leq s_1 \leq 1} \exp\left[ - n\left(\left(H(X_1|X_2) + \frac{L_1}{\sqrt{n}}\right)s_1 \!-\! E_{1}(s_1)\right) \right] \approx \exp\left[ \frac{-L_1^2}{2\sigma^2_{11}} \right],
\end{IEEEeqnarray}
\begin{IEEEeqnarray}{rCl} \label{eq:5-c}
\min_{0 \leq s_3 \leq 1} \exp\left[ - n\left(\left(H(X_1X_2) + \frac{L_1+L_2}{\sqrt{n}}\right)s_1 \!-\! E_{3}(s_3)\right) \right] \approx \exp\left[ \frac{-(L_1+L_2)^2}{2\sigma^2_{33}} \right],
\end{IEEEeqnarray}
for sufficiently large $n$, respectively. The derivation of (\ref{eq:exponent}) and (\ref{eq:5-c}) is as follows: 

In view of (\ref{eq:exponent2}) we have
\begin{IEEEeqnarray*}{rCl}
\lefteqn{\left(H(X_1|X_2) + \frac{L_1}{\sqrt{n}}\right) s_1 \!-\! E_{1}(s_1)} \\
& \approx & \left(H(X_1|X_2) + \frac{L_1}{\sqrt{n}}\right)s_1 - H(X_1|X_2)s_1 - \frac{\sigma_{11}^2s_1^2}{2} \\
& = & \frac{L_1}{\sqrt{n}}s_1 - \frac{\sigma_{11}^2s_1^2}{2}\\
& = & -\frac{\sigma_{11}^{2}}{2} \left( s_1^2 - \frac{2L_1 s_1}{\sqrt{n}{\sigma^2_{11}}} \right) \\
& = & -\frac{\sigma_{11}^{2}}{2} \left( s_1 - \frac{L_1}{\sqrt{n}{\sigma_{11}^2}} \right)^2 + \frac{L_1^2}{2n \sigma^2_{11}},
\end{IEEEeqnarray*}
which is maximized at $s_1 = \frac{L_1}{\sqrt{n}\sigma_{11}^2}$ if $L_1 \geq 0$, so that (\ref{eq:exponent}) follows.
Similarly, in view of (\ref{eq:exponent4}) it follows that
\begin{IEEEeqnarray*}{rCl}
\lefteqn{\left(H(X_1X_2) + \frac{L_1+L_2}{\sqrt{n}}\right) s_3 \!-\! E_{3}(s_3)} \\
& \approx & \left(H(X_1X_2) + \frac{L_1+L_2}{\sqrt{n}}\right)s_3 - H(X_1X_2)s_3 - \frac{\sigma_{33}^2s_3^2}{2} \\
& = & \frac{L_1+L_2}{\sqrt{n}}s_3 - \frac{\sigma_{33}^2s_3^2}{2}\\
& = & -\frac{\sigma_{33}^{2}}{2} \left( s_3 - \frac{L_1 + L_2}{\sqrt{n}{\sigma_{33}^2}} \right)^2 + \frac{\left(L_1+L_2\right)^2}{2n \sigma^2_{33}},
\end{IEEEeqnarray*}
which is maximized at $s_3 = \frac{L_1+L_2}{\sqrt{n}\sigma_{33}^2}$ if $L_1 + L_2 \geq 0$,
so that (\ref{eq:5-c}) holds. 
It should be noted that two maximum values $\frac{L_1^2}{2n\sigma_{11}^2}$ and $\frac{\left( L_1 + L_2 \right)^2}{2n \sigma^2_{33}}$ here could not simultaneously be attained by $s_1$ and $s_3$ if they were constrained so as to be $s_1 = s_3$, as was in the original Koshelev bound \cite{Koshelev} that was used by Tan and Kosut \cite{Tan_ISIT2012}.
On the other hand, in view of (\ref{eq:exponent3}) and $H(X_2) - H(X_2|X_1)>0$, we see that
the second term on the right-hand side of (\ref{gallager1}) is negligible as  $n$ is large enough with $s_2 = \frac{ L_2}{\sqrt{n}\sigma^2_{22}}$.
Thus, in Case I, we have
\begin{IEEEeqnarray}{rCl} \label{gallager2}
\varepsilon_n & \leq & \exp\left[ \frac{-L_1^2}{2\sigma^2_{11}} \right] + \exp\left[ \frac{-(L_1+L_2)^2}{2\sigma^2_{33}} \right] \quad(L_1 \geq 0, L_1+L_2 \geq 0).
\end{IEEEeqnarray} 

In Case II: Similarly to Case I, we have
\begin{IEEEeqnarray}{rCl}  \label{gal:case2}
\varepsilon_n & \leq & \exp\left[ \frac{-(L_1+L_2)^2}{2\sigma^2_{33}} \right] \quad(L_1+L_2 \geq 0) .
\end{IEEEeqnarray} 

In Case III: Similarly to Case I, we have
\begin{IEEEeqnarray}{rCl} \label{gal:case3}
\varepsilon_n & \leq & \exp\left[ \frac{-L_1^2}{2\sigma^2_{11}} \right] \quad(L_1 \geq 0).
\end{IEEEeqnarray} 
%
%
%
%
%
\begin{example} \label{ex1}
We are now ready to give numerical examples. Let us consider i.i.d. correlated sources with binary alphabets ${\cal X}_1 = {\cal X}_2 = \{0,1 \}$, whose joint probabilities are given by Table \ref{table1}.
\begin{table}[htbp] 
\caption{Joint Probability $P_{X_1X_2}(x_1,x_2)$} \label{table1}
\begin{center}
\small
\begin{tabular}{c|c||c|c} 
\multicolumn{2}{c||}{\ } & \multicolumn{2}{c}{$x_1$} \\ \cline{3-4}
\multicolumn{2}{c||}{$P_{X_1X_2}(x_1,x_2)$} & 0 & 1 \\ \hline \hline
\multirow{2}{*}{$x_2$ }& $0$  & $0.5$ & $0.15$ \\ \cline{2-4}
\ & $1$  & $0.25$ & $0.1$ \\ \hline \end{tabular}
\end{center}
\end{table}

The entropies in bits are $H(X_1|X_2) = 0.809$, $H(X_2) =0.934$ and $H(X_1X_2)=1.743$ and the covariance matrix is as follows:
\[
\Sigma_{13} = \left( \begin{array}{cc}
\sigma^2_{11} & \sigma^2_{13} \\
\sigma^2_{31} & \sigma^2_{33} \\
\end{array} \right)
= \left( \begin{array}{cc}
0.475 & 0.492 \\
0.492 & 0.690 \\
\end{array} \right).
\]

In Case I, we compute the error probabilities:
\begin{equation} \label{eq:contour1}
1 - \Phi_{13}(L_1,L_1+L_2)= 1- \int^{L_1}_{-\infty} dy_1 \int^{L_1+L_2}_{-\infty} dy_3 \frac{1}{{2\pi} \sqrt{\det \Sigma_{13}}}  \exp \left( -\frac{1}{2}{\bf y}_{13}\Sigma_{13}^{-1} {\bf y}_{13}^\mathrm{T} \right)
\end{equation}
and 
\begin{IEEEeqnarray}{rCl} \label{eq:contour2}
\exp\left[ \frac{-L_1^2}{2\sigma^2_{11}} \right] + \exp\left[ \frac{-(L_1+L_2)^2}{2\sigma^2_{33}} \right]  \quad (L_1\geq 0, \ L_1+L_2 \geq 0).
\end{IEEEeqnarray} 
Fig. \ref{fig_6} illustrates contour lines of these functions of two variables $L_1$ and $L_2$.
From this result we find that there is a difference between the second-order evaluation and the Koshelev bound, which shows that the former approach outperforms the latter approach. 
On the other hand, Fig. \ref{fig_7} gives an enlargement of the broken contours in Fig. \ref{fig_6}, which is to see the detailed behavior of the curvature for the second-order evaluation.
We observe that in this area the slopes of tangential lines on the broken contours are between $\frac{3\pi}{2}$ and $\frac{7\pi}{4}$. This is an important observation to reveal that the values of $L_1$ are more influential on the error probability than those of $L_2$, but the influence of $L_2$ is not negligible.

In particular, the graphs of functions (\ref{eq:contour1}) and (\ref{eq:contour2}) with $L_1=L_2$ are depicted in Fig. \ref{fig:L_12}.
\begin{figure}[h]
\begin{center}
\includegraphics[width=3.5in]{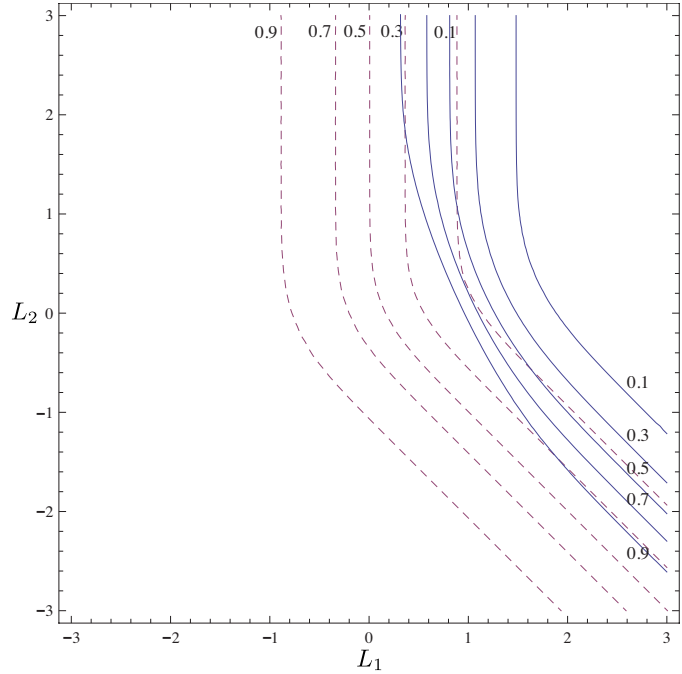}
\centering \caption{Comparison between the Koshelev bound (solid line) and the second-order evaluation (broken line) (Case I): 
contours of functions (\ref{eq:contour1}) and (\ref{eq:contour2}) are drawn. The values of error probabilities are marked.}
\label{fig_6}
\end{center}
\end{figure}
\begin{figure}[h]
\begin{center}
\includegraphics[width=3.5in]{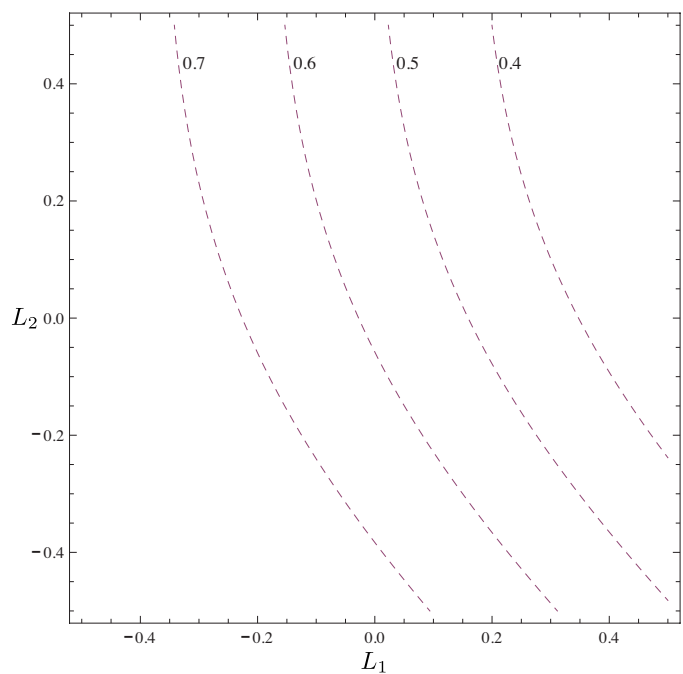}
\centering \caption{Contours of the second-order evaluation. The values $1\!-\!\Phi_{13}(L_1,L_1\!+\!L_2)$ of error probabilities are marked.}
\label{fig_7}
\end{center}
\end{figure}
\begin{figure}[h]
\begin{center}
\includegraphics[width=3.5in]{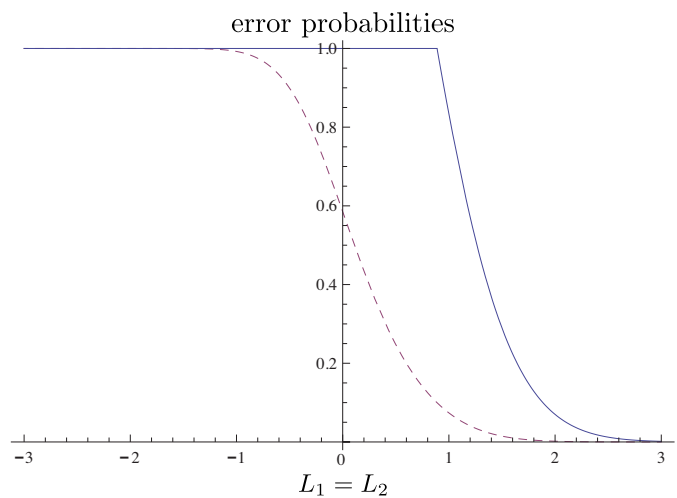}
\centering \caption{The graphs of functions (\ref{eq:contour1}) and (\ref{eq:contour2}) with $L_1 = L_2$ (Case I)}.
\label{fig:L_12}
\end{center}
\end{figure}

\clearpage
In Case II we compute
\[
1 - \Phi_{3} \left( L_1+L_2 \right) = 1 - \int^{L_1+L_2}_{-\infty} \frac{1}{\sqrt{2\pi} \sigma_{33}}  \exp \left( -\frac{y_3^2}{2\sigma^2_{33}} \right) d{y_3},
\]
and
\[
 \exp\left[ \frac{-(L_1+L_2)^2}{2\sigma^2_{33}} \right] \quad (L_1+L_2 \geq0).
\]
Similarly, in Case III we compute
\[
1 - \Phi_{1} \left( L_1\right) = 1-\int^{L_1}_{-\infty} \frac{1}{\sqrt{2\pi} \sigma_{11}}  \exp \left( -\frac{y_1^2}{2\sigma^2_{11}} \right) d{y_1},
\]
and
\[
\exp\left[ \frac{-L_1^2}{2\sigma^2_{11}} \right] \quad (L_1 \geq0).
\]

Fig. \ref{fig_4} and \ref{fig_5} illustrate behaviors of these functions in Case II and Case III, respectively.
In each of these figures, the solid line denotes the Koshelev bound and the broken line denotes the second-order evaluation.
We again find that the second-order evaluation gives {\it better performance} than the Koshelev bound.
%
%
%
%
\begin{figure}[h]
\begin{center}
\includegraphics[width=3.5in]{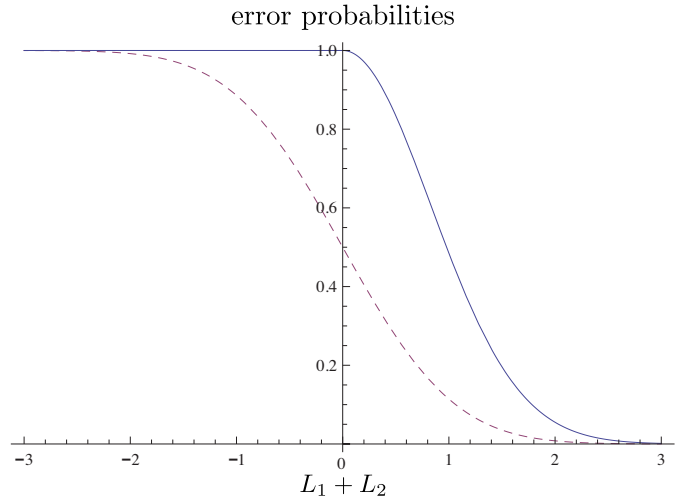}
\centering \caption{Comparison between the Koshelev bound (solid line) and the second-order evaluation (broken line) (Case II)}
\label{fig_4}
\end{center}
\end{figure}
\begin{figure}[h]
\begin{center}
\includegraphics[width=3.5in]{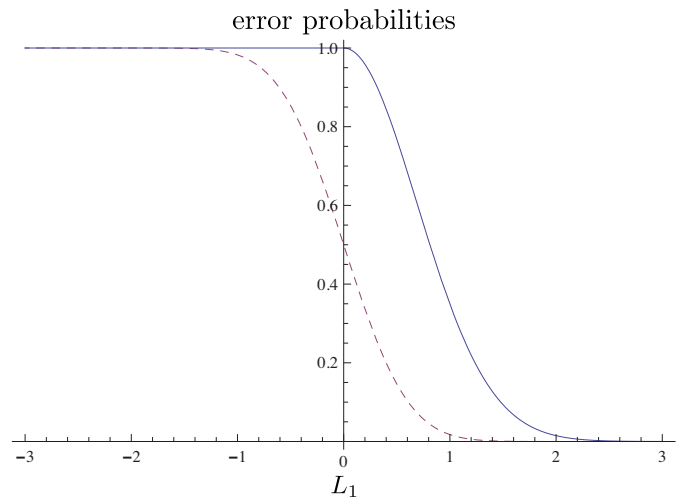}
\centering \caption{Comparison between the Koshelev bound (solid line) and the second-order evaluation (broken line) (Case III)}
\label{fig_5}
\end{center}
\end{figure}
\end{example}
%
%
%
%
%
%
%
%
%
\clearpage
\section{$(a_1, a_2, \varepsilon)$-achievable rate region for Mixed correlated sources} \label{sec:mixed}
In Section \ref{sec:iid}, we have seen that the second-order achievable rate region of any i.i.d. correlated sources (or more generally,  correlated sources whose self-information vector has the multi-dimensional asymptotic normality) is relevantly described on the basis of the information spectrum methods. 
In this section, we establish the second-order achievable rate region of the correlated sources in which the asymptotic normality of self-information vector does {\it not} hold. One of such source classes is the mixed correlated source. Recall that the mixed sources are typical cases of {\it nonergodic} sources.
We consider two typical cases of mixed correlated sources and determine the second-order achievable rate region in each case explicitly by virtue of \lq\lq {\it mixed}" multi-dimensional asymptotic normality.
\subsection{Mixture of Countably Infinite i.i.d. Correlated Sources}
In this subsection, we assume that the correlated sources is a mixture of countably infinite i.i.d. correlated sources.  
Let $\left(X_1^{(k)},X_2^{(k)}\right) \ (k = 1,2,\cdots)$  be arbitrary pairs of i.i.d. correlated sources indexed by $k$.
The  mixed correlated source that we consider in this subsection is defined by
\begin{equation} \label{mix1}
P_{X_1^nX_2^n}({\bf x}_1,{\bf x}_2) = \sum_{k=1}^\infty w(k) P_{X_1^{(k)n}X_2^{(k)n}}({\bf x}_{1},{\bf x}_{2}) ,
\end{equation}
where $\left( {\bf X}_1, {\bf X}_2 \right) = \left\{ \left( X_1^n, X_2^n \right) \right\}_{n=1}^\infty $ and $w(k) \geq 0\ (k=1,2, \cdots)$ are constants such that $\sum_{k=1}^\infty w(k)=1$.

The following lemma, which is implicitly contained in Han \cite{Han}, plays the key role.
We give in this paper the formal proof for the sake of reader's convenience (see Appendix D).
\begin{lemma} \label{lemma1}
Let $\left\{z_n^{(1)} \right\}_{n=1}^\infty$, $\left\{z_n^{(2)} \right\}_{n=1}^\infty$, $\left\{z_n^{(3)} \right\}_{n=1}^\infty$ be any real-valued sequences. Then, for the mixed correlated sources defined by (\ref{mix1}), it holds that, for $k=1,2, \cdots $ with $w(k) >0$;
\begin{enumerate}
\item 
\begin{IEEEeqnarray*}{rCl}
\lefteqn{\Pr\left\{ \frac{-\log P_{X_1^{n}|X_2^{n}}\left( \left. X_1^{(k)n} \right| X_2^{(k)n} \right) }{\sqrt{n}} < z_n^{(1)}, \right.} \\ 
&& \hspace*{0.8cm} \frac{-\log P_{X_2^{n}|X_1^{n}}\left( \left. X_2^{(k)n} \right|X_1^{(k)n}\right) }{\sqrt{n}} < z_n^{(2)},   \\
& & \hspace*{0.8cm} \left. \frac{\!-\!\log P_{X_1^{n}X_2^{n}}\left(X_1^{(k)n}X_2^{(k)n}\right) }{\sqrt{n}} < z_n^{(3)} \right\} \\
& \leq & \Pr\left\{ \frac{\!-\!\log P_{X_1^{(k)n}|X_2^{(k)n}}\left( \left. X_1^{(k)n} \right|X_2^{(k)n}\right) }{\sqrt{n}} < z_n^{(1)}\!+\! \gamma_n, \right. \\ 
&& \hspace*{0.8cm} \frac{-\log P_{X_2^{(k)n}|X_1^{(k)n}}\left( \left. X_2^{(k)n} \right|X_1^{(k)n}\right) }{\sqrt{n}} < z_n^{(2)}\!+\! \gamma_n , \\
& & \hspace*{0.8cm} \left. \frac{-\log P_{X_1^{(k)n}X_2^{(k)n}}\left(X_1^{(k)n}X_2^{(k)n}\right) }{\sqrt{n}} < z_n^{(3)}\!+\!\gamma_n \right\} + 3e^{-\sqrt{n}\gamma_n},
\end{IEEEeqnarray*}
\item 
\begin{IEEEeqnarray*}{rCl}
\lefteqn{ \Pr\left\{ \frac{-\log P_{X_1^{n}|X_2^{n}}\left( \left. X_1^{(k)n}\right|X_2^{(k)n}\right) }{\sqrt{n}} < z_n^{(1)}, \right. } \\
&& \hspace*{0.8cm} \frac{-\log P_{X_2^{n}|X_1^{n}}\left( \left. X_2^{(k)n} \right|X_1^{(k)n}\right) }{\sqrt{n}} < z_n^{(2)}, \\
& &\hspace*{0.8cm}  \left. \frac{\!-\!\log P_{X_1^{n}X_2^{n}}\left(X_1^{(k)n}X_2^{(k)n}\right) }{\sqrt{n}} < z_n^{(3)} \right\} \\
& \geq & \Pr\left\{ \frac{\!-\!\log P_{X_1^{(k)n}|X_2^{(k)n}}\left( \left. X_1^{(k)n}\right|X_2^{(k)n}\right) }{\sqrt{n}} < z_n^{(1)}\!-\! \gamma_n, \right. \\
&& \hspace*{0.8cm}\frac{-\log P_{X_2^{(k)n}|X_1^{(k)n}}\left( \left. X_2^{(k)n} \right|X_1^{(k)n}\right) }{\sqrt{n}} < z_n^{(2)}\!-\! \gamma_n,  \\
& & \hspace*{0.8cm}\left. \frac{-\log P_{X_1^{(k)n}X_2^{(k)n}}\left(X_1^{(k)n}X_2^{(k)n}\right) }{\sqrt{n}} < z_n^{(3)}\!-\!\gamma_n \right\} -  2e^{-\sqrt{n}\gamma_n/2}
\end{IEEEeqnarray*}
\end{enumerate}
for sufficiently large $n$,
where $\gamma_n > 0$ satisfies $\gamma_1 > \gamma_2 > \cdots > 0,$ $\gamma_n \to 0$ and $\sqrt{n} \gamma_n \to \infty$. \QED
\end{lemma}

Now, we define the normal cumulative distribution functions for $k= 1,2,\cdots,$ as follows.
\[
\Phi^{(k)}(T_1,T_2,T_3) \equiv  \int^{T_1}_{-\infty} dy_1 \int^{T_2}_{-\infty} dy_2 \int^{T_3}_{-\infty} d y_3 \frac{1}{(\sqrt{2\pi})^3 \sqrt{\det \Sigma_k}} \exp \left( -\frac{1}{2} {\bf y}  \Sigma_k^{-1} {\bf y}^{\mathrm{T}}  \right),
\]
where ${\bf y}=(y_1,y_2,y_3)$ is a three-dimensional row vector, and $\Sigma_k= (\sigma^2_{ij}(k))\ \ (i,j = 1,2,3; k=1,2, \cdots)$ denotes the covariance matrix, which is given by
\[
\sigma^2_{ij}(k) = \sum_{x_1 \in {\cal X}_1}\sum_{x_2 \in {\cal X}_2} P_{X^{(k)}_1X^{(k)}_2}(x_1,x_2) z_i^{(k)}(x_1,x_2)z_j^{(k)}(x_1,x_2),
\]
where
\[
z_{1}^{(k)}(x_1,x_2) = \log \frac{1}{P_{X^{(k)}_1|X^{(k)}_2}(x_1|x_2)} - H\left(X^{(k)}_1|X^{(k)}_2\right),
\]
\[
z_{2}^{(k)}(x_1,x_2) = \log \frac{1}{P_{X^{(k)}_2|X^{(k)}_1}(x_2|x_1)} - H\left(X^{(k)}_2|X^{(k)}_1\right),
\]
\[
z_{3}^{(k)}(x_1,x_2) = \log \frac{1}{P_{X^{(k)}_1X^{(k)}_2}(x_1,x_2)} - H\left(X^{(k)}_1X^{(k)}_2\right).
\]

In addition, given $a_1, a_2$ and $\varepsilon$, we set
\begin{IEEEeqnarray*}{rCl} 
\lefteqn{K_n^{\mathrm{mix}}(a_1, a_2,\varepsilon|{\bf X}_1,{\bf X}_2)} \nonumber \\
&= & \Biggl\{ (L_1,L_2)  \left| \sum_{k=1}^\infty w(k) \Phi^{(k)} \biggl(\sqrt{n}\left(a_1\!-\! H\left( \left. X_1^{(k)} \right|X_2^{(k)}\right)\right)\!+\! L_1, \right. \nonumber \\ 
&& \hspace*{1.5cm} \sqrt{n}\left(a_2\!-\!H\left( \left. X_2^{(k)} \right|X_1^{(k)}\right) \right) \!+\! L_2, \nonumber \\
& & \hspace*{1.5cm} \sqrt{n}\left(a_1+a_2 - H\left(X_1^{(k)}X_2^{(k)}\right) \right) + L_1+L_2 \biggr) \geq  1 - \varepsilon \Biggr\}.
\end{IEEEeqnarray*}
Then, the following theorem holds.
%
%
%
%
\begin{theorem} \label{theo:mix1}
For the mixed correlated sources defined by (\ref{mix1}), the second-order $(a_1, a_2, \varepsilon)$-achievable rate region is given as the set:
\begin{IEEEeqnarray}{rCl} \label{eq:mix1}
L(a_1, a_2,\varepsilon|{\bf X}_1,{\bf X}_2) &= & \lim_{n \to \infty}K_n^{\mathrm{mix}}(a_1, a_2,\varepsilon|{\bf X}_1,{\bf X}_2) \nonumber \\
& = & \Biggl\{ (L_1,L_2)  \left| \sum_{k=1}^\infty w(k) \lim_{n \to \infty} \Phi^{(k)} \biggl(\sqrt{n}\left(a_1\!-\! H\left( \left. X_1^{(k)} \right|X_2^{(k)}\right)\right)\!+\! L_1, \right. \nonumber \\ 
&& \hspace*{1.5cm} \sqrt{n}\left(a_2\!-\!H\left( \left. X_2^{(k)} \right|X_1^{(k)}\right) \right) \!+\! L_2, \nonumber \\
& & \hspace*{1.5cm} \sqrt{n}\left(a_1+a_2 - H\left(X_1^{(k)}X_2^{(k)}\right) \right) + L_1+L_2 \biggr) \geq  1 - \varepsilon \Biggr\}.
\end{IEEEeqnarray}
 \IEEEQED
\end{theorem} 
This theorem claims that the second-order achievable rate region is given as the set defined by using the  {\it mixed} three-dimensional normal distribution, while in the previous section the set is defined by using the  {\it single} three-dimensional normal distribution. 
Although the quantity in (\ref{eq:mix1}) contains the operation $\lim_{n \to \infty}$, it is possible also to provide an alternative form {\it not} including $\lim_{n \to \infty}$, which is given as a special case of Theorem \ref{theo:mix3} later.
\begin{remark}
It should be noted that, if $a_1 \neq  H\left( \left. X_1^{(k)} \right|X_2^{(k)}\right) \ (k=1,2,\cdots)$, $a_2 \neq  H\left( \left. X_2^{(k)} \right|X_1^{(k)}\right) $ $ (k=1,2,\cdots)$ and $a_1 + a_2 \neq  H\left(X_1^{(k)}X_2^{(k)}\right)$ $(k=1,2,\cdots)$, then it necessarily implies that the second-order achievable rate region is trivial, that is, $L(a_1,a_2,\varepsilon | {\bf X}_1, {\bf X}_2) = \mathbb{R}^2 \mbox{ or } \emptyset$.
On the other hand, it will turn out that in the nontrivial cases the second-order rate region is determined by using a {\it mixture} of two- or one-dimensional normal distributions (cf. Remark \ref{remark:4-2} and Example \ref{example1}).
\end{remark}
%
%
%
%
\begin{IEEEproof}[Proof of Theorem \ref{theo:mix1}]

In view of Theorem \ref{theo:main1}, it suffices to show that
\begin{IEEEeqnarray}{rCl} \label{eq:5-3-0}
\lefteqn{\lim_{n \to \infty} \overline{F}_n(L_1,L_2|a_1,a_2)} \nonumber \\
&= &\sum_{k=1}^\infty w(k) \lim_{n \to \infty}  \Phi^{(k)} \biggl(\sqrt{n}\left(a_1\!-\! H\left( \left. X_1^{(k)} \right|X_2^{(k)}\right)\right)\!+\! L_1, \nonumber \\
&& \hspace*{2.5cm} \sqrt{n}\left(a_2\!-\!H\left( \left. X_2^{(k)}\right|X_1^{(k)}\right) \right) \!+\! L_2,  \nonumber \\
 && \hspace*{2.5cm}  \sqrt{n}\left(a_1+a_2 - H\left(X_1^{(k)}X_2^{(k)}\right) \right) + L_1+L_2 \biggr).
\end{IEEEeqnarray}
By the definition of mixed correlated sources and the first inequality of Lemma \ref{lemma1} with $z_n^{(1)} = \sqrt{n}a_1 + L_1$, $z_n^{(2)} = \sqrt{n}a_2 + L_2$, $z_n^{(3)} = \sqrt{n}\left(a_1 + a_2 \right) + L_1 + L_2$, we have
\begin{IEEEeqnarray*}{rCl}
\lefteqn{\limsup_{n \to \infty} \overline{F}_n(L_1,L_2|a_1,a_2)} \\
& = & \limsup_{n \to \infty} \sum_{k=1}^\infty w(k) \Pr\left\{ \frac{\!-\!\log P_{X_1^{n}|X_2^{n}}\left( \left. X_1^{(k)n} \right|X_2^{(k)n}\right) \!-\!na_1 }{\sqrt{n}}\!<\!L_1, \right. \\
&&  \hspace*{3.5cm} \frac{\!-\!\log P_{X_2^{n}|X_1^{n}}\left(\left. X_2^{(k)n} \right|X_1^{(k)n}\right)\!-\!na_2 }{\sqrt{n}}\!<\!L_2,  \\
& & \hspace*{3.5cm} \left. \frac{-\log P_{X_1^{n}X_2^{n}}\left(X_1^{(k)n}X_2^{(k)n}\right) - n(a_1+a_2) }{\sqrt{n}} < L_1 + L_2 \right\} \\
& \leq & \sum_{k=1}^\infty w(k) \limsup_{n \to \infty} \left( \Pr\left\{ \frac{\!-\!\log P_{X_1^{(k)n}|X_2^{(k)n}}\left( \left. X_1^{(k)n} \right|X_2^{(k)n}\right) -na_1 }{\sqrt{n}} \!<\! L_1\!+\!\gamma_n, \right. \right. \\
&   & \hspace*{3.5cm} \frac{\!-\!\log P_{X_2^{(k)n}|X_1^{(k)n}}\left( \left. X_2^{(k)n}\right|X_1^{(k)n}\right) -na_2 }{\sqrt{n}} \!<\! L_2 \!+\! \gamma_n, \\
& & \hspace*{3.5cm} \left. \left. \frac{-\log P_{X_1^{(k)n}X_2^{(k)n}}\left(X_1^{(k)n}X_2^{(k)n}\right) - n(a_1+a_2)}{\sqrt{n}} < L_1 + L_2 + \gamma_n \right\}  + 3e^{-\sqrt{n}\gamma_n} \right) \\
& = & \sum_{k=1}^\infty w(k) \limsup_{n \to \infty} \Pr\left\{ \frac{\!-\!\log P_{X_1^{(k)n}|X_2^{(k)n}}\left( \left. X_1^{(k)n} \right|X_2^{(k)n}\right) -nH\left(X_1^{(k)n}|X_2^{(k)n}\right) }{\sqrt{n}} \right. \\
&   & \hspace*{6.5cm} \!<\! \sqrt{n}\left(a_1\!-\! H(X_1^{(k)n}|X_2^{(k)n})\right) \!+\!L_1\!+\!\gamma_n, \\
&   & \hspace*{3.5cm} \frac{\!-\!\log P_{X_2^{(k)n}|X_1^{(k)n}}\left( \left. X_2^{(k)n}\right|X_1^{(k)n}\right) -nH(X_2^{(k)n}|X_1^{(k)n}) }{\sqrt{n}} \\
&    & \hspace*{6.5cm} \!<\! \sqrt{n} \left(a_2 -H(X_2^{(k)n}|X_1^{(k)n})  \right) + L_2 \!+\! \gamma_n, \\
& & \hspace*{3.5cm} \frac{-\log P_{X_1^{(k)n}X_2^{(k)n}}\left(X_1^{(k)n}X_2^{(k)n}\right) - nH(X_1^{(k)n}X_2^{(k)n})}{\sqrt{n}} \\
&   & \hspace*{6.5cm} < \sqrt{n}\left( a_1+a_2 - H(X_1^{(k)n}X_2^{(k)n}) \right) + L_1 + L_2 + \gamma_n \Biggr\}.
\end{IEEEeqnarray*}
Then, by virtue of the central limit theorem due to the i.i.d. correlated sources $\left(X_1^{(k)n},X_2^{(k)n} \right) \ (k = 1,2,\cdots)$ as well as the argument similar to the proof of Theorem \ref{main:theo2}, we have
\begin{IEEEeqnarray}{rCl} \label{eq:5-3-1}
\lefteqn{\limsup_{n \to \infty} \overline{F}_n(L_1,L_2|a_1,a_2)} \nonumber \\
&\leq& \sum_{k=1}^\infty w(k) \limsup_{n \to \infty} \Phi^{(k)} \biggl(\sqrt{n}\left(a_1\!-\! H\left(\left. X_1^{(k)} \right|X_2^{(k)}\right)\right)\!+\! L_1, \nonumber \\
&& \hspace*{2.5cm} \sqrt{n}\left(a_2\!-\!H\left(\left.X_2^{(k)}\right|X_1^{(k)}\right) \right) \!+\! L_2,  \nonumber \\
&& \hspace*{2.5cm} \sqrt{n}\left(a_1+a_2 - H\left(X_1^{(k)}X_2^{(k)}\right) \right) + L_1+L_2 \biggr) \nonumber \\
 & = &  \sum_{k=1}^\infty w(k) \lim_{n \to \infty} \Phi^{(k)} \biggl(\sqrt{n}\left(a_1\!-\! H\left( \left. X_1^{(k)} \right|X_2^{(k)}\right)\right)\!+\! L_1, \nonumber \\
&& \hspace*{2.5cm} \sqrt{n}\left(a_2\!-\!H\left( \left. X_2^{(k)} \right|X_1^{(k)}\right) \right) \!+\! L_2,  \nonumber \\
&& \hspace*{2.5cm} \sqrt{n}\left(a_1+a_2 - H\left(X_1^{(k)}X_2^{(k)}\right) \right) + L_1+L_2 \biggr).
\end{IEEEeqnarray}

On the other hand, it can also be verified that, in a manner similar to the above, $\liminf_{n \to \infty} \overline{F}_n(L_1,L_2|a_1,a_2)$ is lower bounded by the right-hand side of (\ref{eq:5-3-1}), that is, 
\begin{IEEEeqnarray}{rCl} \label{eq:5-3-2}
\lefteqn{\liminf_{n \to \infty} \overline{F}_n(L_1,L_2|a_1,a_2)} \nonumber \\
& \geq & \sum_{k=1}^\infty w(k) \lim_{n \to \infty} \Phi^{(k)} \biggl(\sqrt{n}\left(a_1\!-\! H\left( \left. X_1^{(k)} \right|X_2^{(k)}\right)\right)\!+\! L_1, \nonumber \\
&& \hspace*{2.5cm} \sqrt{n}\left(a_2\!-\!H\left( \left. X_2^{(k)} \right|X_1^{(k)}\right) \right) \!+\! L_2,  \nonumber \\
&& \hspace*{2.5cm} \sqrt{n}\left(a_1+a_2 - H\left(X_1^{(k)}X_2^{(k)}\right) \right) + L_1+L_2 \biggr),
\end{IEEEeqnarray}
where the second inequality of Lemma \ref{lemma1} is used instead of the first inequality of Lemma \ref{lemma1}.
Thus, combining (\ref{eq:5-3-1}) and (\ref{eq:5-3-2}) yields (\ref{eq:5-3-0}).

\end{IEEEproof}
\begin{remark}
As shown in the above, the analysis here for mixed correlated sources of i.i.d. sources is based on the asymptotic normality of self-information vector and Lemma \ref{lemma1}. This means that the similar argument is valid for any mixture of countably infinite sources in which the asymptotic normality of self-information vector holds for each of the component correlated sources.
\end{remark}
\begin{example} \label{example1}
Let us consider the mixed correlated sources, for which it holds that $w(1) + w(2) =1$ $(w(1)>0, w(2)>0)$, and
\[
H\left( \left. X_1^{(1)} \right|X_2^{(1)}\right) > H\left(\left. X_1^{(2)}\right|X_2^{(2)}\right),
\]
\[
H\left(X_2^{(1)}\right) > H\left(X_2^{(2)}\right).
\]
We then can compute the second-order achievable rate region:
\[L\left( \left. H\left( \left. X_1^{(1)} \right|X_2^{(1)}\right), H\left(X_2^{(1)}\right),\varepsilon \right|{\bf X}_1,{\bf X}_2\right)\]
and
\[L\left( \left. H\left(\left. X_1^{(2)} \right|X_2^{(2)}\right), H\left(X_2^{(2)}\right),\varepsilon \right |{\bf X}_1,{\bf X}_2\right).\]
Notice that the second-order achievable rate region depends on $w(1)$ and $\varepsilon$. 
From Theorem \ref{theo:mix1} and Remark \ref{remark:4-1}, it is easy to verify that if $w(1) > \varepsilon$ then
\begin{IEEEeqnarray*}{rCl}
\lefteqn{L\left( \left. H\left( \left. X_1^{(1)} \right|X_2^{(1)}\right), H\left(X_2^{(1)}\right),\varepsilon \right|{\bf X}_1,{\bf X}_2\right)} \\
&= & \Biggl\{ (L_1,L_2)  \left| w(1) \Phi_{13}^{(1)} \biggl( L_1, L_1+L_2 \biggr) + w(2) \geq  1 - \varepsilon \right. \Biggr\},
\end{IEEEeqnarray*}
\begin{align*}
L\left( \left. H\left( \left. X_1^{(2)} \right|X_2^{(2)}\right), H\left(X_2^{(2)}\right),\varepsilon \right|{\bf X}_1,{\bf X}_2\right)
= \emptyset,
\end{align*}
whereas if $w(1) < \varepsilon$ then
\begin{align*}
L\left( \left. H\left( \left. X_1^{(1)} \right|X_2^{(1)}\right), H\left(X_2^{(1)}\right),\varepsilon \right|{\bf X}_1,{\bf X}_2\right)
= \mathbb{R}^2,
\end{align*}
\begin{align*}
L\left( \left. H\left( \left. X_1^{(2)} \right|X_2^{(2)}\right), H\left(X_2^{(2)}\right),\varepsilon \right|{\bf X}_1,{\bf X}_2\right)
= \Biggl\{ (L_1,L_2)  \left| w(2) \Phi_{13}^{(2)} \biggl( L_1, L_1+L_2 \biggr) \geq  1 - \varepsilon \right. \Biggr\},
\end{align*}
and otherwise (i.e., $w(1) = \varepsilon$) we have
\begin{align*}
L\left( \left. H\left( \left. X_1^{(1)} \right|X_2^{(1)}\right), H\left(X_2^{(1)}\right),\varepsilon \right|{\bf X}_1,{\bf X}_2\right)
= \mathbb{R}^2,
\end{align*}
\begin{align*}
L\left( \left. H\left( \left. X_1^{(2)} \right|X_2^{(2)}\right), H\left(X_2^{(2)}\right),\varepsilon \right|{\bf X}_1,{\bf X}_2\right)
= \emptyset.
\end{align*}
\end{example}
%
%
%
%
%
%
\subsection{General Mixture of i.i.d. Correlated Sources}
In this subsection, we consider an extension of Theorem \ref{theo:mix1} to the case with {\it general mixture } instead of countably infinite mixtures.
The mixed correlated source $({\bf X}_1, {\bf X}_2)$ that we consider in this subsection is defined by
\begin{equation} \label{mix2}
P_{X_1^nX_2^n}({\bf x}_1,{\bf x}_2) = \int_{\Lambda} P_{X_1^{(\theta)n}X_2^{(\theta)n}}({\bf x}_{1},{\bf x}_{2}) w(d\theta) ,
\end{equation}
where $\left( {\bf X}_1, {\bf X}_2 \right) = \left\{ \left( X_1^n, X_2^n \right) \right\}_{n=1}^\infty $ and $w(d\theta)$ is an {\it arbitrary probability measure} on the parameter space $\Lambda$, and $\left ({\bf X}_1^{(\theta)}, {\bf X}_2^{(\theta)} \right)= \left\{ \left( X_1^{(\theta)n}, X_2^{(\theta)n} \right) \right\}_{n=1}^\infty  \ (\theta \in \Lambda)$ are i.i.d. correlated sources with {\it finite} alphabets, and also the integrand on the right-hand side is assumed to be a measurable function of $\theta$.

We use here the following lemma demonstrated by Han \cite{Han}, instead of Lemma \ref{lemma1}.
\begin{lemma}[Han \cite{Han}] \label{lemma2}
Let $\left\{z_n^{(1)} \right\}_{n=1}^\infty$, $\left\{z_n^{(2)} \right\}_{n=1}^\infty$, $\left\{z_n^{(3)} \right\}_{n=1}^\infty$ be any real-valued sequences. Then, for the mixed correlated sources defined by (\ref{mix2}), it holds that
\begin{IEEEeqnarray*}{rCl}
\lefteqn{\int_{\Lambda} \liminf_{n \to \infty} \Pr\left\{ \frac{\!-\!\log P_{X_1^{(\theta)n}|X_2^{(\theta)n}}\left(X_1^{(\theta)n}\left|X_2^{(\theta)n}\right. \right) }{\sqrt{n}} < z_n^{(1)}\!-\! \gamma_n, \right. }\\
&  & \hspace*{1.5cm} \frac{\!-\!\log P_{X_2^{(\theta)n}|X_1^{(\theta)n}}\left(X_2^{(\theta)n} \left| X_1^{(\theta)n} \right. \right) }{\sqrt{n}} \!<\! z_n^{(2)}\!-\! \gamma_n, \\ 
&& \hspace*{1.5cm} \left. \frac{\!-\!\log P_{X_1^{(\theta)n}X_2^{(\theta)n}}\left(X_1^{(\theta)n}X_2^{(\theta)n}\right) }{\sqrt{n}} \!<\! z_n^{(3)}\!-\!\gamma_n \right\} w(d\theta) \\
& \leq & \liminf_{n \to \infty}\int_{\Lambda} \Pr\left\{ \frac{-\log P_{X_1^{n}|X_2^{n}}\left(X_1^{(\theta)n}\left| X_2^{(\theta)n} \right. \right) }{\sqrt{n}} < z_n^{(1)},  \right. \\
&& \hspace*{1.5cm}  \frac{-\log P_{X_2^{n}|X_1^{n}}\left(X_2^{(\theta)n} \left| X_1^{(\theta)n} \right. \right) }{\sqrt{n}} < z_n^{(2)}, \\
& & \hspace*{1.5cm}\left. \frac{\!-\!\log P_{X_1^{n}X_2^{n}}\left(X_1^{(\theta)n}X_2^{(\theta)n}\right) }{\sqrt{n}} < z_n^{(3)} \right\}w(d\theta) \\
& \leq & \limsup_{n \to \infty}\int_{\Lambda} \Pr\left\{ \frac{-\log P_{X_1^{n}|X_2^{n}}\left(X_1^{(\theta)n}\left| X_2^{(\theta)n} \right. \right) }{\sqrt{n}} < z_n^{(1)}, \right. \\
&& \hspace*{1.5cm}\frac{-\log P_{X_2^{n}|X_1^{n}}\left(X_2^{(\theta)n} \left| X_1^{(\theta)n} \right. \right) }{\sqrt{n}} < z_n^{(2)}, \\
& & \hspace*{1.5cm}\left. \frac{\!-\!\log P_{X_1^{n}X_2^{n}}\left(X_1^{(\theta)n}X_2^{(\theta)n}\right) }{\sqrt{n}} < z_n^{(3)} \right\}w(d\theta) \\
& \leq & \int_{\Lambda} \limsup_{n \to \infty} \Pr\left\{ \frac{\!-\!\log P_{X_1^{(\theta)n}|X_2^{(\theta)n}}\left(X_1^{(\theta)n} \left| X_2^{(\theta)n} \right. \right) }{\sqrt{n}} < z_n^{(1)}\!+\! \gamma_n, \right. \\
&  & \hspace*{1.5cm}\frac{\!-\!\log P_{X_2^{(\theta)n}|X_1^{(\theta)n}}\left(X_2^{(\theta)n} \left| X_1^{(\theta)n} \right. \right) }{\sqrt{n}} \!<\! z_n^{(2)}\!+\! \gamma_n, \\
&& \hspace*{1.5cm} \left. \frac{\!-\!\log P_{X_1^{(\theta)n}X_2^{(\theta)n}}\left(X_1^{(\theta)n}X_2^{(\theta)n}\right) }{\sqrt{n}} \!<\! z_n^{(3)}\!+\!\gamma_n \right\} w(d\theta),
\end{IEEEeqnarray*}
where $\gamma_n > 0$ satisfies $\gamma_1 > \gamma_2 > \cdots > 0$, $\gamma_n \to 0$ and $\sqrt{n} \gamma_n \to \infty$. \IEEEQED
\end{lemma}

In this case, we also define the normal cumulative distribution function for each $\theta \in \Lambda$ as follows.
\[
\Phi^{(\theta)}(T_1,T_2,T_3) \equiv  \int^{T_1}_{-\infty} dy_1 \int^{T_2}_{-\infty} dy_2 \int^{T_3}_{-\infty} d y_3 \frac{1}{(\sqrt{2\pi})^3 \sqrt{\det \Sigma_\theta}} \exp \left( -\frac{1}{2} {\bf y}  \Sigma_\theta^{-1} {\bf y}^\mathrm{T}  \right),
\]
where ${\bf y}=(y_1,y_2,y_3)$ (three-dimensional row vector), and  the covariance matrix $\Sigma_\theta= (\sigma^2_{ij}(\theta))\ \ (i,j = 1,2,3, \theta \in \Lambda)$ are defined in a similar manner to the previous subsection. 

Finally, given $a_1, a_2$ and $\varepsilon$, we set
\begin{multline*}
K_n^{\Lambda}(a_1, a_2,\varepsilon|{\bf X}_1,{\bf X}_2) \\
 \shoveleft{= \Biggl\{ (L_1,L_2)  \left| \int_{\Lambda} \Phi^{(\theta)} \biggl(\sqrt{n}\left(a_1\!-\! H\left(X_1^{(\theta)} \left| X_2^{(\theta)} \right. \right)\right)\!+\! L_1, \sqrt{n}\left(a_2\!-\!H\left(X_2^{(\theta)} \left| X_1^{(\theta)} \right. \right) \right) \!+\! L_2, \right. } \\
   \sqrt{n}\left(a_1+a_2 - H\left(X_1^{(\theta)}X_2^{(\theta)}\right) \right) + L_1+L_2 \biggr)w(d\theta) \geq  1 - \varepsilon \Biggr\}.
\end{multline*}
Then, the following theorem holds.
%
%
%
%
\begin{theorem} \label{theo:mix2}
For the mixed correlated source with {\it finite} alphabet defined by (\ref{mix2}), the second-order $(a_1, a_2, \varepsilon)$-achievable rate region is given as the set:
\begin{IEEEeqnarray}{rCl} \label{eq:mix2}
\lefteqn{L(a_1, a_2,\varepsilon|{\bf X}_1,{\bf X}_2)} \nonumber \\
& = & \lim_{n \to \infty}K_n^{\Lambda}(a_1, a_2,\varepsilon|{\bf X}_1,{\bf X}_2) \nonumber \\
&= & \Biggl\{ (L_1,L_2)  \left| \int_{\Lambda} \lim_{n \to \infty}\Phi^{(\theta)} \biggl(\sqrt{n}\left(a_1\!-\! H\left(X_1^{(\theta)}\left| X_2^{(\theta)} \right.\right)\right)\!+\! L_1, \sqrt{n}\left(a_2\!-\!H\left(X_2^{(\theta)} \left| X_1^{(\theta)} \right. \right) \right) \!+\! L_2, \right. \nonumber \\
&&  \hspace*{2.5cm} \sqrt{n}\left(a_1+a_2 - H\left(X_1^{(\theta)}X_2^{(\theta)}\right) \right) + L_1+L_2 \biggr)w(d\theta) \geq  1 - \varepsilon \Biggr\}.
\end{IEEEeqnarray}
\end{theorem}
\begin{IEEEproof}
It suffices to proceed in parallel with the arguments as made in the proof of Theorem \ref{theo:mix1}. Notice that Lemma \ref{lemma2} (again with $z_n^{(1)}=\sqrt{n} a_1 \!+\! L_1, z_n^{(2)}=\sqrt{n} a_2 \!+\! L_2, z_n^{(3)}=\sqrt{n}(a_1\!+\!a_2) \!+\! L_1\!+\!L_2$)
is used here instead of Lemma \ref{lemma1} 
\end{IEEEproof}

An immediate consequence of Theorem \ref{theo:mix2} is the following compact formula {\it not} including $\lim_{n \to \infty}$, where $\Phi_i^{(\theta)}$, $\Phi_{ij}^{(\theta)}$ are the marginal commutative distribution functions of $\Phi^{(\theta)}$ indicated by $i$, $ij$, respectively (cf. $\Phi_i$, $\Phi_{ij}$ in Section \ref{sec:iid}):
\begin{theorem} \label{theo:mix3}
\begin{IEEEeqnarray}{rCl} \label{eq:mix3}
L(a_1, a_2,\varepsilon|{\bf X}_1,{\bf X}_2) = \left\{ (L_1, L_2)| {\Phi}^\Lambda(a_1, a_2;L_1,L_2) \geq 1 - \varepsilon  \right\},
\end{IEEEeqnarray}
where
\begin{IEEEeqnarray*}{rCl}
\lefteqn{{\Phi}^\Lambda(a_1, a_2;L_1,L_2)} \\
& = & \int_{\Lambda_{\bf 0}(a_1,a_2)} w(d\theta) + \int_{\Lambda_{1}(a_1,a_2)} \Phi_1^{(\theta)}(L_1)w(d\theta) \\
& & + \int_{\Lambda_{ 2}(a_1,a_2)} \Phi_2^{(\theta)}(L_2)w(d\theta) + \int_{\Lambda_{ 3}(a_1,a_2)} \Phi_3^{(\theta)}(L_1+L_2)w(d\theta) \\
& & + \int_{\Lambda_{ 4}(a_1,a_2) \setminus \Lambda_{5}(a_1,a_2)  } \Phi_{13}^{(\theta)}(L_1, L_1+L_2)w(d\theta) + \int_{\Lambda_{5 }(a_1,a_2) \setminus \Lambda_{ 4}(a_1,a_2)  } \Phi_{23}^{(\theta)}(L_2, L_1+L_2)w(d\theta) \\
&& + \int_{\Lambda_{ 4}(a_1,a_2) \cap \Lambda_{5}(a_1,a_2)}  \Phi_{12}^{(\theta)}(L_1, L_2)w(d\theta); 
\end{IEEEeqnarray*}
and 
\begin{IEEEeqnarray*}{rCl}
\Lambda_{ 0}(a_1,a_2) & = & \left\{ \theta \in \Lambda \left| a_1 \!>\! H\left(X_1^{(\theta)}\left| X_2^{(\theta)} \right. \right),\ a_2 \!>\! H\left(X_2^{(\theta)} \left| X_1^{(\theta)} \right. \right),\ a_1\!+\!a_2\!>\!H\left(X^{(\theta)}_1X^{(\theta)}_2 \right)   \right.\right\}, \\
\Lambda_{ 1}(a_1,a_2) & = & \left\{ \theta \in \Lambda \left| a_1 = H\left(X_1^{(\theta)}\left| X_2^{(\theta)} \right. \right),\ a_2 > H\left(X_2^{(\theta)}\right)   \right.\right\}, \\
\Lambda_{ 2}(a_1,a_2) & = & \left\{ \theta \in \Lambda \left| a_1 > H\left(X_1^{(\theta)} \right),\ a_2 = H\left(X_2^{(\theta)} \left| X_1^{(\theta)} \right. \right) \right.\right\}, \\
\Lambda_{3}(a_1,a_2) & = & \left\{ \theta \in \Lambda \left| a_1 \!>\! H\left(X_1^{(\theta)}\left| X_2^{(\theta)} \right. \right),\ a_2 \!>\! H\left(X_2^{(\theta)} \left| X_1^{(\theta)} \right. \right),\ a_1\!+\!a_2\!=\!H\left(X^{(\theta)}_1X^{(\theta)}_2 \right)   \right.\right\}, \\
\Lambda_{4}(a_1,a_2) & = & \left\{ \theta \in \Lambda \left| a_1 = H\left(X_1^{(\theta)}\left| X_2^{(\theta)} \right. \right),\ a_2 = H\left(X_2^{(\theta)} \right) \right.\right\}, \\
\Lambda_{5}(a_1,a_2) & = & \left\{ \theta \in \Lambda \left| a_1 = H\left(X_1^{(\theta)} \right),\ a_2 = H\left(X_2^{(\theta)} \left| X_1^{(\theta)} \right. \right) \right.\right\}.
\end{IEEEeqnarray*}
\end{theorem}
\begin{IEEEproof}
It suffices to scrutinize in Theorem \ref{theo:mix2} the situation with $n \to \infty$ and take account of Theorem \ref{main:theo2}.
\end{IEEEproof}
\begin{remark}
It is easily seen that if $X_1^{(\theta)}$ and $X_2^{(\theta)}$ are correlated for all $\theta \in \Lambda$ then all $\Lambda_i(a_1,a_2) \quad (i=0,1,\dots, 5)$ are mutually disjoint, which, in particular, implies that
\begin{IEEEeqnarray*}{rCl}
\Lambda_4(a_1,a_2)\setminus \Lambda_5(a_1,a_2) & = & \Lambda_4(a_1,a_2) \\
\Lambda_5(a_1,a_2)\setminus \Lambda_4(a_1,a_2) & = & \Lambda_5(a_1,a_2) \\
\Lambda_4(a_1,a_2) \cap \Lambda_5(a_1,a_2) & = & \emptyset.
\end{IEEEeqnarray*}
On the other hand, if $X_1^{(\theta)}$ and $X_2^{(\theta)}$ are independent for some $\theta \in \Lambda$ then 
$\Lambda_4(a_1,a_2) \cap \Lambda_5(a_1,a_2) \neq \emptyset$; in this case it should be noted that if $\theta \in \Lambda_4(a_1,a_2) \cap \Lambda_5(a_1,a_2)$ then 
\begin{IEEEeqnarray*}{rCl}
\Phi_{13}^{(\theta)}(L_1, L_1+L_2)=\Phi_{23}^{(\theta)}(L_2, L_1+L_2)=\Phi_{12}^{(\theta)}(L_1,L_2).
\end{IEEEeqnarray*}
\IEEEQED
\end{remark}
\section{Concluding Remarks} \label{sec:con}
So far we have established the {\it second-order} source coding theorem for the Slepian-Wolf system with {\it general} correlated sources. 
On the other hand, in the single-user source coding problem, Hayashi \cite{Hayashi} has shown the optimal {\it second-order} achievable rate for {\it general} sources, and actually computed it for an i.i.d. source using the asymptotic normality. 
In the case that the source is assumed to be in a class of {\it mixed} sources, Nomura and Han \cite{NH2011} have demonstrated the  {\it second-order} achievable rates explicitly.

Analogously, with the Slepian-Wolf coding system, we have established the {\it second-order} achievable rate region for {\it general} correlated sources, and actually computed it for i.i.d. correlated sources and mixed correlated sources, respectively.
As we have mentioned in the above, in order to compute the achievable rate region for i.i.d. correlated sources, the property of multivariate normal distribution played the key role. In particular, the condition for the {\it second-order} achievable rate region turned out to be expressed in terms of the marginal distributions of a three-dimensional normal distribution. 

In order to elucidate the effectiveness of Theorem \ref{main:theo2} we have numerically compared it with a slight strengthening of the original Koshelev bound, and it turned out that the former outperforms the latter.

Notice here that in \cite{NH2011} the optimal second-order rates for mixed sources in the single-user source coding problem has also been 
expressed in terms of the {\it single} equation such as (\ref{eq:mix1}) and (\ref{eq:mix2}), which we called there the {\it canonical representation}, introduced for the first time in this paper.
This is indeed a great advantage of the information spectrum methods.
While the equation in \cite{NH2011} is based on the single-normal distribution functions, the similar formulas in the present paper are based on the three-dimensional normal distribution functions.
From this viewpoint, the results in the present paper can be considered as a reasonable generalization of the results in \cite{NH2011}. 
In the channel coding context, Polyanskiy, Poor and Verd\'{u} \cite{Polyanskiy2011} have derived the optimal second-order capacity for the Gilbert-Elliott channel by using a form similar to canonical equations.

Finally, it should be emphasized that our analysis is based on the information spectrum methods and hence the results in this paper are valid with {\it countably infinite} alphabet excepting Theorem \ref{theo:mix2} with {\it general mixtures}.
%
\appendices
%
%
%
%
%
\section{Proof of Theorem \ref{theo:main1}}
\renewcommand{\theequation}{A.\arabic{equation}}
\setcounter{equation}{0}

It is obvious by virtue of De Morgan's law that (\ref{fn}) in Theorem \ref{theo:main1} is equivalent to 
\begin{align} \label{fn2}
L(a_1, a_2,\varepsilon|{\bf X}_1,{\bf X}_2) = \mbox{Cl} \Bigl( \Bigl\{ (L_1,L_2) & \left| \liminf_{n \to \infty} \overline{F}_n(L_1,L_2|a_1,a_2) \geq  1 - \varepsilon \right\} \Bigr),
\end{align}
so that in the sequel we will give the proof of (\ref{fn2}) instead of (\ref{fn}).
 \\
1) {\it Direct Part: } 

 For any fixed $(L_1,L_2)$ satisfying 
 \begin{equation} \label{eq:3-1-0}
 \liminf_{n \to \infty} \overline{F}_n(L_1,L_2|a_1,a_2) \geq 1 -\epsilon,
 \end{equation}
 set
 \[
 M^{(1)}_n = e^{n a_1 + L_1\sqrt{n} +  2\sqrt[4]{n}\gamma}, \ \ 
 M^{(2)}_n = e^{n a_2 + L_2\sqrt{n} +  2\sqrt[4]{n}\gamma},
 \]
 where $\gamma >0$ is an arbitrary small constant.
It is obvious that
 \[
\limsup_{n \to \infty} \frac{1}{\sqrt{n}} \log \frac{M_n^{(1)}}{e^{na_1}} \leq L_1,
\]
and
\[
\limsup_{n \to \infty} \frac{1}{\sqrt{n}} \log \frac{M_n^{(2)}}{e^{na_2}} \leq L_2
\]
hold.
Thus, in this direct part it suffices to show the existence of an $(n, M_n^{(1)},M_n^{(2)}, \epsilon_n)$ code such that $\limsup_{n \to \infty} \varepsilon_n \leq \varepsilon$.

Lemma \ref{lemma1-1} implies that there exists an $(n, M_n^{(1)},M_n^{(2)}, \epsilon_n)$ code such that
\begin{align*}
\varepsilon_n  \leq & \Pr \left\{ z_n P_{X_1^n|X_2^n}(X_1^n|X_2^n) \leq e^{- n a_1 - L_1\sqrt{n} - 2\sqrt[4]{n}\gamma} \right. \\
& \hspace*{0.5cm} \mbox{ or } z_n P_{X_2^n|X_1^n}(X_2^n|X_1^n) \leq e^{- n a_2 - L_2\sqrt{n} - 2\sqrt[4]{n}\gamma} \\
& \hspace*{0.5cm} \left.  \mbox{or } z_n P_{X_1^nX_2^n}(X_1^n,X_2^n) \leq e^{- n( a_1 + a_2) - \sqrt{n}(L_1 + L_2) - 4\sqrt[4]{n}\gamma} \right\} + 3z_n \\
 = &  \Pr \left\{ \frac{1}{\sqrt{n}} \log \frac{1}{z_n P_{X_1^n|X_2^n}(X_1^n|X_2^n)} \geq \frac{n a_1 + L_1\sqrt{n} + 2 \sqrt[4]{n}\gamma}{\sqrt{n}}, \right. \\
   & \ \ \ \ \  \mbox{   or }  \frac{1}{\sqrt{n}} \log \frac{1}{z_n P_{X_2^n|X_1^n}(X_2^n|X_1^n)} \geq \frac{n a_2 + L_2\sqrt{n} + 2 \sqrt[4]{n}\gamma}{\sqrt{n}}, \\
  & \ \ \ \ \ \left.\mbox{or }  \frac{1}{\sqrt{n}} \log \frac{1}{z_n P_{X_1^nX_2^n}(X_1^nX_2^n)} \geq \frac{n (a_1 + a_2) + (L_1+L_2)\sqrt{n} + 4 \sqrt[4]{n}\gamma}{\sqrt{n}} \right\} + 3z_n.
\end{align*}
Since $\{z_n\}_{n=1}^\infty$ is a sequence of arbitrary numbers satisfying $z_i >0$ $(\forall i=1,2,\cdots)$, we set
$z_n = e^{- \sqrt[4]{n} \gamma }$. 
Then, we have
\begin{align*} 
\varepsilon_n \leq & \Pr \left\{ \frac{1}{\sqrt{n}} \log \frac{1}{P_{X_1^n|X_2^n}(X_1^n|X_2^n)} \geq \frac{n a_1 + L_1\sqrt{n} + 2\sqrt[4]{n} \gamma}{\sqrt{n}} - \frac{\sqrt[4]{n}\gamma}{\sqrt{n}}, \right. \nonumber \\
&  \ \ \ \ \  \mbox{   or }  \frac{1}{\sqrt{n}} \log \frac{1}{ P_{X_2^n|X_1^n}(X_2^n|X_1^n)} \geq \frac{n a_2 + L_2\sqrt{n} + 2\sqrt[4]{n}\gamma}{\sqrt{n}} - \frac{\sqrt[4]{n}\gamma}{\sqrt{n}}, \nonumber \\
&   \ \ \ \ \ \left.\mbox{or }  \frac{1}{\sqrt{n}} \log \frac{1}{ P_{X_1^nX_2^n}(X_1^nX_2^n)} \geq \frac{n (a_1 + a_2) + (L_1+L_2)\sqrt{n} + 4\sqrt[4]{n}\gamma}{\sqrt{n}} -\frac{\sqrt[4]{n}\gamma}{ \sqrt{n}} \right\} + 3e^{-\sqrt[4]{n} \gamma }.
\end{align*}
Hence,
\begin{eqnarray} \label{eq:3-1-1}
\varepsilon_n & \leq & \Pr \left\{ \frac{-\log P_{X_1^n|X_2^n}(X_1^n|X_2^n) - n a_1}{\sqrt{n}} \geq L_1 + \frac{ \gamma}{\sqrt[4]{n}}, \right. \nonumber \\
&   & \ \ \ \ \  \mbox{   or }  \frac{-\log P_{X_2^n|X_1^n}(X_2^n|X_1^n) - n a_2}{\sqrt{n}} \geq  L_2 + \frac{\gamma}{\sqrt[4]{n}}, \nonumber \\
&   & \ \ \ \ \ \left.\mbox{or }  \frac{-\log P_{X_1^nX_2^n}(X_1^nX_2^n) - n( a_1+a_2)}{\sqrt{n}} \geq L_1+L_2 + \frac{3\gamma}{\sqrt[4]{n}} \right\} + 3e^{- \sqrt[4]{n}\gamma } \nonumber \\
& \leq & \Pr \left\{ \frac{-\log P_{X_1^n|X_2^n}(X_1^n|X_2^n) - n a_1}{\sqrt{n}} \geq L_1 + \frac{ \gamma}{\sqrt[4]{n}}, \right. \nonumber \\
&   & \ \ \ \ \  \mbox{   or }  \frac{-\log P_{X_2^n|X_1^n}(X_2^n|X_1^n) - n a_2}{\sqrt{n}} \geq  L_2 + \frac{\gamma}{\sqrt[4]{n}}, \nonumber \\
&   & \ \ \ \ \ \left.\mbox{or }  \frac{-\log P_{X_1^nX_2^n}(X_1^nX_2^n) - n( a_1+a_2)}{\sqrt{n}} \geq L_1+L_2 + \frac{2\gamma}{\sqrt[4]{n}} \right\} + 3e^{- \sqrt[4]{n} \gamma }.
\end{eqnarray}
Hence, by means of De Morgan's law, we have
\begin{align} \label{morgan}
\Pr &\left\{ \frac{\!-\log P_{X_1^n|X_2^n}(X_1^n|X_2^n) - n a_1}{\sqrt{n}} \geq L_1 + \frac{\gamma}{\sqrt[4]{n}}, \right. 
\nonumber \\
& \ \ \ \ \ \mbox{  or }  \frac{\!-\log P_{X_2^n|X_1^n}(X_2^n|X_1^n) - n a_2}{\sqrt{n}} \geq  L_2 + \frac{\gamma}{\sqrt[4]{n}}, \nonumber \\
& \ \ \ \ \    \left.\mbox{or }  \frac{-\log P_{X_1^nX_2^n}(X_1^nX_2^n) - n( a_1+a_2)}{\sqrt{n}} \geq L_1+L_2 +\frac{2\gamma}{\sqrt[4]{n}} \right\} \nonumber  \\
 =  &1\!-\!\Pr \left\{ \frac{\!-\!\log P_{X_1^n|X_2^n}(X_1^n|X_2^n) - n a_1}{\sqrt{n}} < L_1\!+\!\frac{\gamma}{\sqrt[4]{n}}, \right. \nonumber \\
 & \ \ \ \ \ \ \ \frac{\!-\log P_{X_2^n|X_1^n}(X_2^n|X_1^n) - n a_2}{\sqrt{n}} <  L_2\!+\!\frac{\gamma}{\sqrt[4]{n}}, \nonumber \\
&  \ \ \ \ \ \ \   \left. \frac{-\log P_{X_1^nX_2^n}(X_1^nX_2^n) - n( a_1\!+\!a_2)}{\sqrt{n}} < L_1+L_2  + \frac{2\gamma}{\sqrt[4]{n}} \right\} \nonumber \\
= &  1-\overline{F}_n \left( \left. L_1 + \frac{\gamma}{\sqrt[4]{n}} ,L_2 +\frac{\gamma}{\sqrt[4]{n}} \right|a_1,a_2 \right).
\end{align}
Substituting (\ref{morgan}) into (\ref{eq:3-1-1}), we have
\begin{eqnarray*}
\varepsilon_n & \leq & 1-\overline{F}_n \left( \left. L_1 + \frac{\gamma}{\sqrt[4]{n}} ,L_2 +\frac{\gamma}{\sqrt[4]{n}} \right|a_1,a_2 \right) + 3e^{- \sqrt[4]{n}\gamma} \\
&  \leq & 1- \overline{F}_n \left( \left. L_1 ,L_2 \right|a_1,a_2 \right) + 3e^{-\sqrt[4]{n}\gamma}.
\end{eqnarray*}
By taking $\limsup_{n \to \infty}$ of both sides, we have
\begin{eqnarray*}
\limsup_{n \to \infty} \varepsilon_n & \leq & 1- \liminf_{n \to \infty} \overline{F}_n \left( \left. L_1 ,L_2 \right|a_1,a_2 \right) \leq \epsilon,
\end{eqnarray*}
where the last inequality follows from (\ref{eq:3-1-0}). 
Thus, the direct part has been proved.
\vspace*{\baselineskip}
\\
2) {\it Converse Part:}

Suppose that a pair $(L_1,L_2)$
is $(a_1, a_2, \varepsilon)$-achievable. Then, from the assumption, for any small $\gamma >0$ we have a code $(n, M_n^{(1)}, M_n^{(2)}, \varepsilon_n)$ such that  
 \begin{equation} \label{eq:3-1-2}
 \frac{1}{\sqrt{n}} \log \frac{M_n^{(1)}}{e^{na_1}} \leq L_1 + \gamma,
\end{equation}
\begin{equation} \label{eq:3-1-3}
\frac{1}{\sqrt{n}} \log \frac{M_n^{(2)}}{e^{na_2}} \leq L_2 + \gamma, 
\end{equation}
for all sufficiently large $n$, and that
\begin{equation} \label{eq:3-1-5}
\limsup_{n \to \infty} \varepsilon_n \leq \varepsilon.
\end{equation}

Thus, substituting (\ref{eq:3-1-2}) and (\ref{eq:3-1-3}) into Lemma \ref{lemma1-2}, 
the error probability is lower bounded by
\begin{eqnarray*}
\varepsilon_n & \geq & \Pr \left\{ \frac{-\log P_{X_1^n|X_2^n}(X_1^n|X_2^n) - na_1}{\sqrt{n}} \geq L_1 + \gamma -\frac{\log z_n}{\sqrt{n}}\right. \\
&& \ \ \ \ \  \mbox{or } \frac{-\log P_{X_2^n|X_1^n}(X_2^n|X_1^n) - na_2}{\sqrt{n}} \geq L_2 + \gamma -\frac{\log z_n}{\sqrt{n}}\\
&   & \ \ \ \ \ \left.\mbox{or } \frac{-\log P_{X_1^nX_2^n}(X_1^nX_2^n) - n(a_1 + a_2)}{\sqrt{n}} \geq L_1 + L_2 + 2 \gamma -\frac{\log z_n}{\sqrt{n}} \right\} - 3z_n,
\end{eqnarray*}
for all $n=1,2,\cdots,$ where $\{z_n\}_{n=1}^\infty$ is a sequence of an arbitrary numbers satisfying $z_i >0 \quad (\forall i =1,2,\cdots)$. 
Set $z_n = e^{- \sqrt{n} \gamma}$ and substituting it into the above, we have
\begin{eqnarray*}
\varepsilon_n & \geq & \Pr \left\{ \frac{-\log P_{X_1^n|X_2^n}(X_1^n|X_2^n) - na_1}{\sqrt{n}} \geq L_1  +2 \gamma \right. \\
&& \ \ \ \ \  \mbox{or } \frac{-\log P_{X_2^n|X_1^n}(X_2^n|X_1^n) - na_2}{\sqrt{n}} \geq L_2 +2 \gamma \\
&   & \ \ \ \ \ \left.\mbox{or } \frac{-\log P_{X_1^nX_2^n}(X_1^nX_2^n) - n(a_1 + a_2)}{\sqrt{n}} \geq L_1 + L_2 +3 \gamma \right\} - 3e^{-\sqrt{n}\gamma} \\
& \geq & \Pr \left\{ \frac{-\log P_{X_1^n|X_2^n}(X_1^n|X_2^n) - na_1}{\sqrt{n}} \geq L_1  +2 \gamma \right. \\
&& \ \ \ \ \  \mbox{or } \frac{-\log P_{X_2^n|X_1^n}(X_2^n|X_1^n) - na_2}{\sqrt{n}} \geq L_2 +2 \gamma \\
&   & \ \ \ \ \ \left.\mbox{or } \frac{-\log P_{X_1^nX_2^n}(X_1^nX_2^n) - n(a_1 + a_2)}{\sqrt{n}} \geq L_1 + L_2 + 4 \gamma \right\} - 3e^{-\sqrt{n}\gamma}
\end{eqnarray*}
Here, again, owing to De Morgan's law, we have
\begin{eqnarray*}
\varepsilon_n & \geq & 1 - \Pr \left\{ \frac{-\log P_{X_1^n|X_2^n}(X_1^n|X_2^n) - na_1}{\sqrt{n}} < L_1 +2 \gamma,\right. \\
&& \ \ \ \ \  \frac{-\log P_{X_2^n|X_1^n}(X_2^n|X_1^n) - na_2}{\sqrt{n}} < L_2 +2 \gamma,  \\
&   &\ \ \ \ \  \left. \frac{-\log P_{X_1^nX_2^n}(X_1^nX_2^n) - n(a_1 + a_2)}{\sqrt{n}} < L_1 + L_2 +4  \gamma \right\} - 3e^{-\sqrt{n}\gamma} \\
& = & 1- \overline{F}_n \left( \left. L_1 +2 \gamma ,L_2 +2 \gamma \right|a_1,a_2 \right) - 3e^{-\sqrt{n} \gamma}.
\end{eqnarray*}
By taking $\limsup_{n \to \infty}$ in (\ref{eq:3-1-5}) we have
\begin{align*}
\varepsilon \geq \limsup_{n \to \infty} \varepsilon_n & \geq 1- \liminf_{n \to \infty} \overline{F}_n \left( \left. L_1 +2\gamma ,L_2  +2\gamma \right|a_1,a_2 \right).
\end{align*}
Since $\gamma >0$ is arbitrary, this means that 
\begin{align*}
L(a_1, a_2,\varepsilon|{\bf X}_1,{\bf X}_2) \subset \mbox{Cl} \Bigl( \Bigl\{ (L_1,L_2) & \left| \liminf_{n \to \infty} \overline{F}_n(L_1,L_2|a_1,a_2) \geq  1 - \varepsilon \right\} \Bigr).
\end{align*}
Thus, we have proved the converse part.
\IEEEQED
%
%
%
%
%
%
\newpage
\section{Proof of Theorem \ref{main:theo2}}
\renewcommand{\theequation}{B.\arabic{equation}}
\setcounter{equation}{0}
{\it 1) Case I:}

In view of Theorem \ref{theo:main1} in the form of (\ref{fn2}), it suffices to calculate
\begin{IEEEeqnarray*}{rCl}
\lefteqn{\liminf_{n \to \infty} \overline{F}_n(L_1,L_2|a_1,a_2)} \\
& = & \liminf_{n \to \infty} \Pr \left\{ \frac{-\log P_{X^n_1|X^n_2}(X_1^n|X_2^n) - na_1} {\sqrt{n}} < L_1, \right. \\
& & \hspace*{2cm}\frac{-\log P_{X^n_2|X^n_1}(X_2^n|X_1^n) - na_2} {\sqrt{n}} < L_2,  \\
& & \hspace*{2cm}\left. \frac{-\log P_{X^n_1X^n_2}(X_1^nX_2^n) - n \left(a_1 + a_2\right)} {\sqrt{n}} < L_1 + L_2 \right\}.
\end{IEEEeqnarray*}
Substituting $a_1 = H(X_1|X_2)$ and $a_2 = H(X_2)$ into $\overline{F}_n(L_1,L_2|a_1,a_2)$, we have
\begin{IEEEeqnarray*}{rCl}
\lefteqn{\overline{F}_n(L_1,L_2|a_1,a_2)} \\ 
&=& \Pr \left\{ \frac{-\log P_{X^n_1|X^n_2}(X_1^n|X_2^n) - nH(X_1|X_2)} {\sqrt{n}} < L_1, \right. \\
&& \hspace*{0.8cm} \frac{-\log P_{X^n_2|X^n_1}(X_2^n|X_1^n) - nH(X_2)} {\sqrt{n}} < L_2,   \\
&& \hspace*{0.8cm} \left. \frac{-\log P_{X^n_1X^n_2}(X_1^nX_2^n) - n \left(H(X_1|X_2) + H(X_2)\right)} {\sqrt{n}} < L_1 + L_2 \right\} \\
&= &\Pr \left\{ \frac{-\log P_{X^n_1|X^n_2}(X_1^n|X_2^n) - nH(X_1|X_2)} {\sqrt{n}} < L_1, \right. \\
&&\hspace*{0.8cm}  \frac{-\log P_{X^n_2|X^n_1}(X_2^n|X_1^n) - nH(X_2)} {\sqrt{n}} < L_2,  \\
& &\hspace*{0.8cm} \left. \frac{-\log P_{X^n_1X^n_2}(X_1^nX_2^n) - n H(X_1X_2)} {\sqrt{n}} < L_1 + L_2 \right\} \\
&= &\Pr \left\{ \frac{-\log P_{X^n_1|X^n_2}(X_1^n|X_2^n) - nH(X_1|X_2)} {\sqrt{n}} < L_1, \right. \\
&& \frac{-\log P_{X^n_2|X^n_1}(X_2^n|X_1^n) - nH(X_2|X_1) } {\sqrt{n}} < \sqrt{n}\left (H(X_2) - H(X_2|X_1) \right) + L_2, \\
&& \left. \frac{-\log P_{X^n_1X^n_2}(X_1^nX_2^n) - n H(X_1X_2)} {\sqrt{n}} < L_1 + L_2 \right\},
\end{IEEEeqnarray*}
where the second equality is derived by the chain rule $H(X_1|X_2) + H(X_2) = H(X_1X_2)$.
Moreover, $H(X_2) - H(X_2|X_1) = I(X_1;X_2) > 0$ holds, since we are considering correlated sources (recall that we have assumed that $\Sigma$ is positive-definite; also cf. the observation below). Thus, for any constant $W >0$, we have $\sqrt{n}\left(H(X_2) - H(X_2|X_1) \right) > W$ for sufficiently large $n$.

As a consequence, noting that the correlated sources has an i.i.d. property, and taking account of the asymptotic normality (due to the central limit theorem), we have
\begin{align*}
\overline{F}_n(L_1,L_2|a_1,a_2) \geq & \Pr \left\{ \frac{\!-\! \log P_{X^n_1|X^n_2}(X_1^n|X_2^n)\!-\!nH(X_1|X_2)} {\sqrt{n}}\!<\!L_1, \right.  \\
& \ \ \ \ \ \frac{\!-\! \log P_{X^n_2|X^n_1}(X_2^n|X_1^n)\!-\!nH(X_2|X_1) } {\sqrt{n}}\!< W+L_2, \\
& \ \ \ \ \ \left. \frac{\!-\log P_{X^n_1X^n_2}(X_1^nX_2^n) - n H(X_1X_2)} {\sqrt{n}}\!<\!L_1 + L_2 \right\} \\
\to & \Phi \left( L_1,W+L_2, L_1+L_2\right) \ \ (\mbox{as }n \to \infty).
\end{align*}
Furthermore, as $W > 0$ can be arbitrarily large, this implies that
\begin{align} \label{g1}
\liminf_{n \to \infty} \overline{F}_n(L_1,L_2|a_1,a_2) & \geq \lim_{W \to \infty} \Phi \left( L_1,W+L_2, L_1+L_2\right)  = \Phi_{13} \left( L_1, L_1+L_2 \right). 
\end{align}
On the other hand, it is obvious that 
\begin{align*}
 \overline{F}_n(L_1,L_2|a_1,a_2) 
\leq & \Pr \left\{ \frac{-\log P_{X^n_1|X^n_2}(X_1^n|X_2^n) - nH(X_1|X_2)} {\sqrt{n}} < L_1, \right. \\
& \ \ \ \ \  \left. \frac{-\log P_{X^n_1X^n_2}(X_1^nX_2^n) - n H(X_1X_2)} {\sqrt{n}} < L_1 + L_2 \right\},
\end{align*}
so that we have
\begin{align} \label{g2}
\limsup_{n \to \infty} \overline{F}_n(L_1,L_2|a_1,a_2) & \leq \Phi_{13} \left( L_1, L_1+L_2 \right).
\end{align}
Thus, summarizing (\ref{g1}) and (\ref{g2}) leads to
\begin{align}  \label{eq:g3}
\lim_{n \to \infty} \overline{F}_n(L_1,L_2|a_1,a_2) & = \Phi_{13} \left( L_1, L_1+L_2 \right). 
\end{align}
Therefore, the proof of Case I has been completed  with the observation that in the case of $I(X_1;X_2)=0$ the right-hand side of (\ref{eq:g3}) turns out to coincide with $\Phi_{12}(L_1, L_2)$.
\vspace*{\baselineskip}\\
{\it 2) Case II:}

First, notice that 
\begin{eqnarray*}
a_1+ a_2 & = & \lambda H(X_1) + (1-\lambda) H(X_1|X_2) + (1-\lambda) H(X_2) + \lambda H(X_2|X_1) \\
& = & \lambda\left( H(X_1) + H(X_2|X_1) \right) + ( 1-  \lambda ) \left( H(X_1|X_2) + H(X_2) \right) \\
& = & H(X_1X_2).
\end{eqnarray*}
Moreover, $\Delta_1 \equiv a_1 - H(X_1|X_2) >0$ because $X_1$ and $X_2$ are correlated and
\[
a_1 - H(X_1|X_2) = \lambda H(X_1) + (1-\lambda) H(X_1|X_2) -H(X_1|X_2) = \lambda \left( H(X_1) - H(X_1|X_2)  \right) > 0.
\]
The same argument yields $\Delta_2 \equiv a_2 - H(X_2|X_1) > 0$, so that
$\overline{F}_n(L_1,L_2|a_1,a_2)$ is given by
\begin{align*}
\overline{F}_n(L_1,L_2|a_1,a_2) 
= \Pr &\left\{ \frac{-\log P_{X^n_1|X^n_2}(X_1^n|X_2^n) - nH(X_1|X_2)} {\sqrt{n}} < \Delta_1\sqrt{n} + L_1, \right. \\
& \frac{-\log P_{X^n_2|X^n_1}(X_2^n|X_1^n) - nH(X_2|X_1) } {\sqrt{n}} < \Delta_2 \sqrt{n} + L_2, \\
& \left. \frac{-\log P_{X^n_1X^n_2}(X_1^nX_2^n) - n H(X_1X_2)} {\sqrt{n}} < L_1 + L_2 \right\}.
\end{align*}
Here, ${\Delta_1\sqrt{n}} $ and ${\Delta_2\sqrt{n}}$ goes to $\infty$ as $n \to \infty$.
Therefore, again, by virtue of the asymptotic normality it holds that
\begin{align*}
\lim_{n \to \infty} \overline{F}_n(L_1,L_2|a_1,a_2) = \Phi_{3}(L_1 + L_2).
\end{align*}
Therefore, the proof of Case II has been completed. 
\vspace*{\baselineskip}\\
{\it 3) Case III:}

Notice that setting $\delta = a_2 - H(X_2) > 0$ leads to
\begin{eqnarray*}
a_1+ a_2 - H(X_1X_2) & = & H(X_1|X_2) + H(X_2)  + \delta  - H(X_1X_2)\\
& = & \delta > 0.
\end{eqnarray*}
Moreover, $\Delta_2 = a_2 - H(X_2|X_1) >0$ holds, because
\[
a_2 - H(X_2|X_1) = H(X_2) + \delta - H(X_2|X_1)  > 0.
\]
holds.
Thus,
$\overline{F}_n(L_1,L_2|a_1,a_2)$ is given by
\begin{align*}
\overline{F}_n(L_1,L_2|a_1,a_2) 
= \Pr &\left\{ \frac{-\log P_{X^n_1|X^n_2}(X_1^n|X_2^n) - nH(X_1|X_2)} {\sqrt{n}} < L_1, \right. \\
& \ \ \ \ \ \frac{-\log P_{X^n_2|X^n_1}(X_2^n|X_1^n) - nH(X_2|X_1) } {\sqrt{n}} < \Delta_2 \sqrt{n} + L_2, \\
& \ \ \ \ \ \left. \frac{-\log P_{X^n_1X^n_2}(X_1^nX_2^n) - n H(X_1X_2)} {\sqrt{n}} < \delta \sqrt{n} +L_1 + L_2 \right\},
\end{align*}
Here, ${\Delta_2\sqrt{n}} $ and ${\delta \sqrt{n}}$ goes to $\infty$ as $n \to \infty$.
Then, again, by virtue of the asymptotic normality as well as the same discussion in Cases I and II,  it is concluded that
\begin{align*}
\lim_{n \to \infty} \overline{F}_n(L_1,L_2|a_1,a_2) = \Phi_{1}(L_1),
\end{align*}
thus completing the proof.
\IEEEQED
%
%
%
%
%
%
\section{Proof of Lemma \ref{exponent}}
\renewcommand{\theequation}{C.\arabic{equation}}
\setcounter{equation}{0}
We slightly modify the Koshelev's argument \cite{Koshelev}. We first define the encoders $\phi_n^{(1)}, \phi_n^{(2)}$ and the decoder $\psi_n$ as follows.

{\it Encoder:}

For each source output ${\bf x}_1 \in {\cal X}_1^n$ we randomly generate an index $i_1 \in {\cal M}_n^{(1)} \equiv \left\{ 1,2, \cdots, M_n^{(1)} \right\} $ according to the uniform distribution and set $\phi_n^{(1)}({\bf x}_1) = i_1$. The encoder $\phi_n^{(2)}({\bf x}_2) = i_2 \in {\cal M}_n^{(2)} \equiv \left\{ 1,2, \cdots, M_n^{(2)} \right\}  $ is defined similarly.

{\it Decoder:}

Suppose that a decoder $\psi_n$ receives a pair of the encoder outputs $(i_1, i_2)$. 
We define the maximum likelihood decoder 
\[
\psi_n(i_1, i_2) = \arg \max_{({\bf x}_1, {\bf x}_2) \in {\cal X}_1^n \times {\cal X}_2^n}Q_{i_1,i_2}({\bf x}_1,{\bf x}_2) 
\]
where
\[
Q_{i_1,i_2}({\bf x}_1,{\bf x}_2) = P_{X_1^nX_2^n}({\bf x}_1,{\bf x}_2) {\mathbbm 1}\left\{ \phi_n^{(1)}({\bf x}_1) = i_1,\phi_n^{(2)}({\bf x}_2) = i_2 \right\},
\]
and ${\mathbbm 1}\{ \cdot \}$ denotes the indicator function.

{\it Analysis of error probability:}

The error probability $\varepsilon_n$ is evaluated as follows where $\Pr \{ \cdot \}$ denotes the probability due to the random code:
\begin{IEEEeqnarray*}{rCl}
{\varepsilon}_n & \leq & \sum_{({\bf x}_1, {\bf x}_2) \in {\cal X}_1^n \times {\cal X}_2^n}P_{X_1^nX_2^n}({\bf x}_1, {\bf x}_2)  \Pr \left\{ \bigcup_{({\bf x}'_1,{\bf x}'_2) \neq ({\bf x}_1,{\bf x}_2)}\left\{\frac{Q_{\phi_n^{(1)}({\bf x}_1),\phi_n^{(2)}({\bf x}_2)  }({\bf x}'_1,{\bf x}'_2)}{P_{X_1^nX_2^n}({\bf x}_1,{\bf x}_2)} \geq 1 \right\} \right\} \\
& \leq & \sum_{({\bf x}_1, {\bf x}_2) \in {\cal X}_1^n \times {\cal X}_2^n} P_{X_1^nX_2^n}({\bf x}_1, {\bf x}_2) \Pr \left\{ \bigcup_{{\bf x}'_1 \neq {\bf x}_1}\left\{ \frac{Q_{\phi_n^{(1)}({\bf x}_1),\phi_n^{(2)}({\bf x}_2)}({\bf x}'_1, {\bf x}_2) }{P_{X_1^nX_2^n}({\bf x}_1, {\bf x}_2)} \geq 1 \right\} \right\}  \nonumber \\ 
&& + \sum_{({\bf x}_1, {\bf x}_2) \in {\cal X}_1^n \times {\cal X}_2^n}P_{X_1^nX_2^n}({\bf x}_1, {\bf x}_2)  \Pr \left\{ \bigcup_{ {\bf x}'_2 \neq {\bf x}_2} \left\{ \frac{Q_{\phi_n^{(1)}({\bf x}_1),\phi_n^{(2)}({\bf x}_2)}({\bf x}_1, {\bf x}'_2)}{P_{X_1^nX_2^n}({\bf x}_1, {\bf x}_2)} \geq 1 \right\} \right\}  \nonumber \\ 
&& + \sum_{({\bf x}_1, {\bf x}_2) \in {\cal X}_1^n \times {\cal X}_2^n}P_{X_1^nX_2^n}({\bf x}_1, {\bf x}_2) \Pr \left\{ \bigcup_{{\bf x}'_1 \neq{\bf x}_1, {\bf x}'_2 \neq{\bf x}_2}\left\{ \frac{Q_{\phi_n^{(1)}({\bf x}_1),\phi_n^{(2)}({\bf x}_2)}({\bf x}'_1, {\bf x}'_2)}{P_{X_1^nX_2^n}({\bf x}_1, {\bf x}_2)} \geq 1 \right\} \right\}.
\end{IEEEeqnarray*}
Then, with any $t_1, t_2, t_3 \geq 0$ and any $0 \leq s_1, s_2, s_3 \leq 1$, we obtain
\begin{IEEEeqnarray*}{rCl} \label{app4-1}
{\varepsilon_n} & \leq & \sum_{({\bf x}_1, {\bf x}_2) \in {\cal X}_1^n \times {\cal X}_2^n}P_{X_1^nX_2^n}({\bf x}_1,{\bf x}_2) \left[ \sum_{ {\bf x}'_1 \neq {\bf x}_1} {\mathbb E}\left(\frac{Q_{\phi_n^{(1)}({\bf x}_1),\phi_n^{(2)}({\bf x}_2)  }({\bf x}'_1,{\bf x}_2)}{P_{X_1^nX_2^n}({\bf x}_1,{\bf x}_2)} \right)^{t_1} \right]^{s_1} \nonumber \\
&& + \sum_{({\bf x}_1, {\bf x}_2) \in {\cal X}_1^n \times {\cal X}_2^n}P_{X_1^nX_2^n}({\bf x}_1,{\bf x}_2) \left[ \sum_{{\bf x}'_2 \neq {\bf x}_2} {\mathbb E}\left(\frac{Q_{\phi_n^{(1)}({\bf x}_1),\phi_n^{(2)}({\bf x}_2)  }({\bf x}_1,{\bf x}'_2)}{P_{X_1^nX_2^n}({\bf x}_1,{\bf x}_2)} \right)^{t_2} \right]^{s_2} \nonumber \\
&& + \sum_{({\bf x}_1, {\bf x}_2) \in {\cal X}_1^n \times {\cal X}_2^n}P_{X_1^nX_2^n}({\bf x}_1,{\bf x}_2) \left[ \sum_{ {\bf x}'_1 \neq {\bf x}_1,{\bf x}'_2 \neq {\bf x}_2} {\mathbb E} \left(\frac{Q_{\phi_n^{(1)}({\bf x}_1),\phi_n^{(2)}({\bf x}_2)  }({\bf x}'_1,{\bf x}'_2)}{P_{X_1^nX_2^n}({\bf x}_1,{\bf x}_2)}\right)^{t_3} \right]^{s_3},
\end{IEEEeqnarray*}
where ${\mathbb E}$ denotes the expectation due to the random code, from which it follows that
\begin{IEEEeqnarray}{rCl} \label{app4-2}
{\varepsilon_n} & \leq & \sum_{({\bf x}_1, {\bf x}_2) \in {\cal X}_1^n \times {\cal X}_2^n}P_{X_1^nX_2^n}({\bf x}_1,{\bf x}_2)^{1-t_1s_1} \left( \sum_{ {\bf x}'_1 \neq {\bf x}_1} {\mathbb E} Q_{\phi_n^{(1)}({\bf x}_1),\phi_n^{(2)}({\bf x}_2)  }({\bf x}'_1,{\bf x}_2)^{t_1} \right)^{s_1} \nonumber \\
&& + \sum_{({\bf x}_1, {\bf x}_2) \in {\cal X}_1^n \times {\cal X}_2^n}P_{X_1^nX_2^n}({\bf x}_1,{\bf x}_2)^{1-t_2s_2} \left( \sum_{{\bf x}'_2 \neq {\bf x}_2} {\mathbb E}Q_{\phi_n^{(1)}({\bf x}_1),\phi_n^{(2)}({\bf x}_2)  }({\bf x}_1,{\bf x}'_2)^{t_2} \right)^{s_2} \nonumber \\
&& + \sum_{({\bf x}_1, {\bf x}_2) \in {\cal X}_1^n \times {\cal X}_2^n}P_{X_1^nX_2^n}({\bf x}_1,{\bf x}_2)^{1-t_3s_3} \left( \sum_{ {\bf x}'_1 \neq {\bf x}_1,{\bf x}'_2 \neq {\bf x}_2} {\mathbb E}Q_{\phi_n^{(1)}({\bf x}_1),\phi_n^{(2)}({\bf x}_2)  }({\bf x}'_1,{\bf x}'_2)^{t_3} \right)^{s_3}. \nonumber \\
&&
\end{IEEEeqnarray}
Here, the first term of the right-hand side of the above inequality is evaluated as follows with $t_1 = \frac{1}{1+s_1}$: 
\begin{IEEEeqnarray*}{rCl}
\lefteqn{ \sum_{({\bf x}_1, {\bf x}_2) \in {\cal X}_1^n \times {\cal X}_2^n}P_{X_1^nX_2^n}({\bf x}_1,{\bf x}_2)^{\frac{1}{1+s_1}} \left( \sum_{ {\bf x}'_1 \neq {\bf x}_1} {\mathbb E}Q_{\phi_n^{(1)}({\bf x}_1),\phi_n^{(2)}({\bf x}_2)  }({\bf x}'_1,{\bf x}_2)^{\frac{1}{1+s_1}} \right)^{s_1}} \\
& = & \sum_{({\bf x}_1, {\bf x}_2) \in {\cal X}_1^n \times {\cal X}_2^n}P_{X_1^nX_2^n}({\bf x}_1,{\bf x}_2)^{\frac{1}{1+s_1}} \left( \sum_{ {\bf x}'_1 \neq {\bf x}_1} P_{X_1^nX_2^n}({\bf x}'_1,{\bf x}_2)^{\frac{1}{1+s_1}} {\mathbb E} {\mathbbm 1}\left\{ \phi_n^{(1)}({\bf x}_1) = \phi_n^{(1)}({\bf x}'_1) \right\} \right)^{s_1} \\
& = & \left( \frac{1}{M_n^{(1)}} \right)^{s_1}\sum_{({\bf x}_1, {\bf x}_2) \in {\cal X}_1^n \times {\cal X}_2^n}P_{X_1^nX_2^n}({\bf x}_1,{\bf x}_2)^{\frac{1}{1+s_1}} \left( \sum_{ {\bf x}'_1 \neq {\bf x}_1} P_{X_1^nX_2^n}({\bf x}'_1,{\bf x}_2)^{\frac{1}{1+s_1}} \right)^{s_1} \\
& \leq & \left( \frac{1}{M_n^{(1)}} \right)^{s_1}\sum_{ {\bf x}_2 \in  {\cal X}_2^n}\left( \sum_{ {\bf x}_1 \in {\cal X}_1^n} P_{X_1^nX_2^n}({\bf x}_1,{\bf x}_2)^{\frac{1}{1+s_1}} \right)^{1+s_1} .
\end{IEEEeqnarray*}
The second and the third terms on the right-hand side of (\ref{app4-2}) are similarly bounded from above by setting $t_2 = \frac{1}{1+s_2}$, $t_3 = \frac{1}{1+s_3}$, respectively.

Summarizing the above arguments, it is concluded that the error probability $\varepsilon_n$ is upper bounded by
\begin{IEEEeqnarray*}{rCl}
{\varepsilon}_n & \leq & \left( \frac{1}{M_n^{(1)}} \right)^{s_1}\sum_{ {\bf x}_2 \in  {\cal X}_2^n}\left( \sum_{ {\bf x}_1 \in {\cal X}_1^n} P_{X_1^nX_2^n}({\bf x}_1,{\bf x}_2)^{\frac{1}{1+s_1}} \right)^{1+s_1} \\
&& + \left( \frac{1}{M_n^{(2)}} \right)^{s_2}\sum_{ {\bf x}_1 \in  {\cal X}_1^n}\left( \sum_{ {\bf x}_2 \in {\cal X}_2^n} P_{X_1^nX_2^n}({\bf x}_1,{\bf x}_2)^{\frac{1}{1+s_2}} \right)^{1+s_2} \\
&& + \left( \frac{1}{M_n^{(1)}M_n^{(2)}} \right)^{s_3}\left( \sum_{ ({\bf x}_1,{\bf x}_2) \in {\cal X}_1^n \times {\cal X}_2^n} P_{X_1^nX_2^n}({\bf x}_1,{\bf x}_2)^{\frac{1}{1+s_3}} \right)^{1+s_3},
\end{IEEEeqnarray*}
thus completing the proof of the lemma, because we are considering stationary memoryless correlated sources (cf. Gallager \cite{Gallager}).

%
%
%
%
%
%
\section{Proof of Lemma \ref{lemma1}}
\renewcommand{\theequation}{D.\arabic{equation}}
\setcounter{equation}{0}
We first show the first inequality. 
Set a sequence $\{\gamma_n \}_{n=1}^{\infty}$  satisfying $\gamma_1 > \gamma_2 > \cdots > 0,$ and $\gamma_n \to 0$, $\sqrt{n} \gamma_n \to \infty$ and define three sets 
\begin{align*}
D_n^{(1)}(k) = & \left\{({\bf x}_1,{\bf x}_2) \in {\cal X}_1^n \times {\cal X}_2^n \left| \frac{-\!\log P_{X_1^n|X_2^n}({\bf x}_1|{\bf x}_2) }{\sqrt{n}}\!-\!\frac{-\!\log P_{X^{(k)n}_{1}|X^{(k)n}_{2}}({\bf x}_1|{\bf x}_2) }{\sqrt{n}}\!\leq - \gamma_n \right. \right\},
\end{align*}
\begin{align*}
D_n^{(2)}(k) = & \left\{({\bf x}_1,{\bf x}_2) \in {\cal X}_2^n \times {\cal X}_1^n \left| \frac{-\!\log P_{X_2^n|X_1^n}({\bf x}_2|{\bf x}_1) }{\sqrt{n}}\!-\!\frac{-\!\log P_{X^{(k)n}_{2}|X^{(k)n}_{1}}({\bf x}_2|{\bf x}_1) }{\sqrt{n}}\!\leq - \gamma_n \right. \right\},
\end{align*}
\begin{align*}
D_n^{(3)}(k) = & \left\{({\bf x}_1,{\bf x}_2) \in {\cal X}_1^n \times {\cal X}_2^n \left| \frac{-\!\log P_{X_1^nX_2^n}({\bf x}_1,{\bf x}_2) }{\sqrt{n}}\!-\!\frac{-\!\log P_{X^{(k)n}_{1}X^{(k)n}_{2}}({\bf x}_1,{\bf x}_2) }{\sqrt{n}}\!\leq - \gamma_n \right. \right\}
\end{align*}
for $k = 1,2, \cdots$.
In addition, we set
\begin{align*}
D_n(k) = D_n^{(1)}(k) \cup D_n^{(2)}(k) \cup D_n^{(3)}(k).
\end{align*}
Then, it holds that for $k=1,2,\cdots$
\begin{align} \label{eq:app1-0}
\Pr\left\{  X_{1}^{(k)n}X_{2}^{(k)n} \in D_n^{(3)}(k) \right\} & = \sum_{({\bf x}_1, {\bf x}_2) \in D_n^{(3)}(k)} P_{X_{1}^{(k)n}X_{2}^{(k)n}}({\bf x}_1, {\bf x}_2) \nonumber \\
& \leq \sum_{({\bf x}_1, {\bf x}_2) \in D_n^{(3)}(k)} P_{X_{1}^{n}X_{2}^{n}}({\bf x}_1,{\bf x}_2) e^{-\sqrt{n}\gamma_n} \nonumber \\
& \leq e^{-\sqrt{n} \gamma_n}.
\end{align}
Similarly, it holds that
\begin{align} \label{eq:app1-1}
\Pr\left\{ X_{1}^{(k)n}X_{2}^{(k)n} \in D_n^{(1)}(k) \right\} & = \sum_{({\bf x}_1, {\bf x}_2) \in D_n^{(1)}(k)} P_{X_{1}^{(k)n}X_{2}^{(k)n}}({\bf x}_1, {\bf x}_2) \nonumber \\
& =  \sum_{({\bf x}_1, {\bf x}_2) \in D_n^{(1)}(k)} P_{X_{1}^{(k)n}|X_{2}^{(k)n}}({\bf x}_1| {\bf x}_2)P_{X_{2}^{(k)n}}({\bf x}_2) \nonumber \\
& \leq \sum_{({\bf x}_1, {\bf x}_2) \in D_n^{(1)}(k)} P_{X_{1}^{n}|X_{2}^{n}}({\bf x}_1|{\bf x}_2)P_{X_{2}^{(k)n}}({ \bf x}_2) e^{-\sqrt{n}\gamma_n} \nonumber \\
& \leq e^{-\sqrt{n} \gamma_n},
\end{align}
for $k=1,2,\cdots$ and also that
\begin{align} \label{eq:app1-3}
\Pr\left\{ X_{1}^{(k)n}X_{2}^{(k)n} \in D_n^{(2)}(k) \right\} & \leq e^{-\sqrt{n}\gamma_n}
\end{align}
for $k = 1,2,\cdots$.
It then follows from (\ref{eq:app1-0}), (\ref{eq:app1-1}) and (\ref{eq:app1-3}) that
\begin{align*}
\Pr\left\{ X_{1}^{(k)n},X_{2}^{(k)n} \in D_n(k) \right\}  \leq 3e^{-\sqrt{n}\gamma_n}.
\end{align*}
This means that
\begin{align*}
\lefteqn{\Pr\left\{ X_{1}^{(k)n}X_{2}^{(k)n} \notin D_n(k) \right\}} \\
& =  \Pr\left\{ \frac{-\log P_{X_1^{n}|X_2^{n}}\left(X_1^{(k)n}|X_2^{(k)n}\right) }{\sqrt{n}} > \frac{-\log P_{X_1^{(k)n}|X_2^{(k)n}}\left(X_1^{(k)n}|X_2^{(k)n}\right) }{\sqrt{n}} - \gamma_n,  \right. \\
&  \hspace*{2.5cm} \frac{-\log P_{X_2^{n}|X_1^{n}}\left(X_2^{(k)n}|X_1^{(k)n}\right) }{\sqrt{n}} > \frac{-\log P_{X_2^{(k)n}|X_1^{(k)n}}\left(X_2^{(k)n}|X_1^{(k)n}\right) }{\sqrt{n}} - \gamma_n, \\
&  \hspace*{2.5cm}\left. \frac{-\log P_{X_1^{n}X_2^{n}}\left(X_1^{(k)n}X_2^{(k)n}\right) }{\sqrt{n}} > \frac{-\log P_{X_1^{(k)n}X_2^{(k)n}}\left(X_1^{(k)n}X_2^{(k)n}\right) }{\sqrt{n}} - \gamma_n \right\} \\
&  \geq 1 - 3e^{-\sqrt{n}\gamma_n}.
\end{align*}
Hence, we have
\begin{IEEEeqnarray*}{rCl}
\lefteqn{ \Pr\left\{ \frac{-\log P_{X_1^{(k)n}|X_2^{(k)n}}\left(X_1^{(k)n}|X_2^{(k)n}\right) }{\sqrt{n}} -\gamma_n < z_n^{(1)},  \right. } \\
&& \hspace*{0.8cm}\frac{-\log P_{X_2^{(k)n}|X_1^{(k)n}}\left(X_2^{(k)n}|X_1^{(k)n}\right) }{\sqrt{n}} -\gamma_n < z_n^{(2)}, \\
& & \hspace*{0.8cm} \left. \frac{\!-\!\log P_{X_1^{(k)n}X_2^{(k)n}}\left(X_1^{(k)n}X_2^{(k)n}\right) }{\sqrt{n}} - \gamma_n < z_n^{(3)} \right\} \\
& \geq & \Pr\left\{ \frac{\!-\!\log P_{X_1^{n}|X_2^{n}}\left(X_1^{(k)n}|X_2^{(k)n}\right) }{\sqrt{n}} < z_n^{(1)}, \right. \\
&&\hspace*{0.8cm} \frac{-\log P_{X_2^{n}|X_1^{n}}\left(X_2^{(k)n}|X_1^{(k)n}\right) }{\sqrt{n}} < z_n^{(2)},   \\
& &\hspace*{0.8cm} \left. \frac{-\log P_{X_1^{n}X_2^{n}}\left(X_1^{(k)n}X_2^{(k)n}\right) }{\sqrt{n}} < z_n^{(3)} \right\} - 3e^{-\sqrt{n}\gamma_n},
\end{IEEEeqnarray*}
for $ k = 1,2,\cdots$, which is the first inequality of the lemma.

Next, we show the second inequality of the lemma.
Set 
\begin{equation*}
S_n^{(1)} = \left\{ ({\bf x}_1,{\bf x}_2) \in {\cal X}_1^n \times {\cal X}_2^n \left| \log P_{X_1^{n}|X_2^{n}}({\bf x}_1|{\bf x}_2) \geq - \sqrt{n} z_n^{(1)} \right. \right\},
\end{equation*}
\begin{equation*}
S_n^{(2)} = \left\{ ({\bf x}_1,{\bf x}_2) \in {\cal X}_1^n \times {\cal X}_2^n \left| \log P_{X_2^{n}|X_1^{n}}({\bf x}_2|{\bf x}_1) \geq - \sqrt{n} z_n^{(2)} \right. \right\},
\end{equation*}
\begin{equation*}
S_n^{(3)} = \left\{ ({\bf x}_1,{\bf x}_2) \in {\cal X}_1^n \times {\cal X}_2^n \left| \log P_{X_1^{n}X_2^{n}}({\bf x}_1,{\bf x}_2) \geq - \sqrt{n} z_n^{(3)} \right. \right\}.
\end{equation*}
In the sequel, setting $S_n = S_n^{(1)} \cap S_n^{(2)} \cap S_n^{(3)}$, we evaluate $P_{X_1^{(k)n}X_2^{(k)n}}(S_n)$.
To do so, first rewrite  $\log P_{X_1^{n}|X_2^{n}}({\bf x}_1|{\bf x}_2)$ as 
\begin{IEEEeqnarray}{rCl} \label{delta1}
\log P_{X_1^{n}|X_2^{n}}({\bf x}_1|{\bf x}_2) & = & \log P_{X_1^{n}X_2^{n}}({\bf x}_1,{\bf x}_2) - \log P_{X_2^{n}}({\bf x}_2) \nonumber \\
& = & \log \sum_{k=1}^\infty w(k) P_{X_1^{(k)n}X_2^{(k)n}}({\bf x}_1,{\bf x}_2) - \log P_{X_2^{n}}({\bf x}_2) \nonumber \\
& \geq & \log w(k) + \log P_{X_1^{(k)n}X_2^{(k)n}}({\bf x}_1,{\bf x}_2) - \log P_{X_2^{(k)n}}({\bf x}_2)  \nonumber \\
&& + \left( \log P_{X_2^{(k)n}}({\bf x}_2) - \log P_{X_2^{n}}({\bf x}_2) \right) \nonumber \\
& \geq & -\sqrt{n} \gamma_n + \log P_{X_1^{(k)n}|X_2^{(k)n}}({\bf x}_1|{\bf x}_2) \nonumber \\
&&+ \left( \log P_{X_2^{(k)n}}({\bf x}_2) - \log P_{X_2^{n}}({\bf x}_2) \right) \nonumber \\
& \equiv & \delta^{(k)}_1({\bf x}_1, {\bf x}_2),
\end{IEEEeqnarray}
where we have taken account of $\gamma_n \geq \frac{- \log w(k)}{\sqrt{n}}$ for $k$ with $w(k) >0$ and sufficiently large $n$.

Similarly, we have
\begin{IEEEeqnarray}{rCl} \label{delta2}
\log P_{X_2^{n}|X_1^{n}}({\bf x}_2|{\bf x}_1) & \geq & -\sqrt{n} \gamma_n + \log P_{X_2^{(k)n}|X_1^{(k)n}}({\bf x}_2|{\bf x}_1) \nonumber \\
&&+ \left( \log P_{X_1^{(k)n}}({\bf x}_1) - \log P_{X_1^{n}}({\bf x}_1) \right) \nonumber \\
& \equiv & \delta^{(k)}_2({\bf x}_1, {\bf x}_2),
\end{IEEEeqnarray}
\begin{IEEEeqnarray}{rCl} \label{delta3}
\log P_{X_1^{n}X_2^{n}}({\bf x}_1,{\bf x}_2) & \geq & -\sqrt{n} \gamma_n + \log P_{X_1^{(k)n}X_2^{(k)n}}({\bf x}_1,{\bf x}_2) \nonumber \\
& \equiv & \delta^{(k)}_3({\bf x}_1, {\bf x}_2).
\end{IEEEeqnarray}

On the other hand, again similarly to (\ref{eq:app1-0}), it follows that
\begin{IEEEeqnarray*}{rCl}
\Pr\left\{ X_1^{(k)n} \in E_n^{(1)} \right\} \geq 1 - e^{-\sqrt{n}\gamma_n},
\end{IEEEeqnarray*}
\begin{IEEEeqnarray*}{rCl}
\Pr\left\{ X_2^{(k)n} \in E_n^{(2)} \right\} \geq 1 - e^{-\sqrt{n}\gamma_n},
\end{IEEEeqnarray*}
where
\begin{IEEEeqnarray*}{rCl}
E_n^{(1)} = \left\{ {\bf x}_1 \in {\cal X}^n_1 \left| - \log P_{X_1^n}({\bf x}_1) +  \log P_{X_1^{(k)n}}({\bf x}_1) \geq -\sqrt{n}\gamma_n \right. \right\},
\end{IEEEeqnarray*}
\begin{IEEEeqnarray*}{rCl}
E_n^{(2)} = \left\{ {\bf x}_2 \in {\cal X}^n_2 \left| - \log P_{X_2^n}({\bf x}_2) +  \log P_{X_2^{(k)n}}({\bf x}_2) \geq -\sqrt{n}\gamma_n \right. \right\}.
\end{IEEEeqnarray*}

Hence, with $E_n = E_n^{(1)} \times E_n^{(2)} $  we obtain
\begin{IEEEeqnarray}{rCl} \label{eq:appc-3}
\Pr \left\{ X_1^{(k)n}X_2^{(k)n} \in E_n \right\} \geq 1 - 2e^{-\sqrt{n}\gamma_n}.
\end{IEEEeqnarray}

Let us now define the subsets $G_n^{(1)}(k)$, $G_n^{(2)}(k)$ and $G_n^{(3)}(k)$ by
\begin{IEEEeqnarray}{rCl} \label{G1}
G_n^{(1)}(k) = \left\{ ({\bf x}_1,{\bf x}_2) \in {\cal X}_1^n \times {\cal X}_2^n \left| \delta_1^{(k)}({\bf x}_1,{\bf x}_2) \geq - \sqrt{n} z_n^{(1)} \right. \right\},
\end{IEEEeqnarray}
\begin{IEEEeqnarray}{rCl} \label{G2}
G_n^{(2)}(k) = \left\{ ({\bf x}_1,{\bf x}_2) \in {\cal X}_1^n \times {\cal X}_2^n \left| \delta_2^{(k)}({\bf x}_1,{\bf x}_2) \geq - \sqrt{n} z_n^{(2)} \right. \right\},
\end{IEEEeqnarray}
\begin{IEEEeqnarray}{rCl} \label{G3}
G_n^{(3)}(k) = \left\{ ({\bf x}_1,{\bf x}_2) \in {\cal X}_1^n \times {\cal X}_2^n \left| \delta_3^{(k)}({\bf x}_1,{\bf x}_2) \geq - \sqrt{n} z_n^{(3)} \right. \right\}.
\end{IEEEeqnarray}

It is then obvious that, for $k=1,2, \cdots, $
\begin{IEEEeqnarray*}{rCl}
S_n^{(1)} \supset G_n^{(1)}(k), \ \ S_n^{(2)} \supset G_n^{(2)}(k), \ \ S_n^{(3)} \supset G_n^{(3)}(k).
\end{IEEEeqnarray*}

Thus, setting $G_n{(k)} = G_n^{(1)}{(k)} \cap G_n^{(2)}{(k)} \cap G_n^{(3)}{(k)}$, we obtain
\begin{IEEEeqnarray}{rCl} \label{eq:appc-0}
P_{X_1^{(k)n}X_2^{(k)n}}(S_n) & \geq & P_{X_1^{(k)n}X_2^{(k)n}}(G_n(k)) \nonumber \\
& \geq & P_{X_1^{(k)n}X_2^{(k)n}}(G_n(k) \cap E_n) \nonumber \\
& = & P_{X_1^{(k)n}X_2^{(k)n}}(G_n(k) |E_n)P_{X_1^{(k)n}X_2^{(k)n}}(E_n).
\end{IEEEeqnarray}

Now we set 
\begin{IEEEeqnarray*}{rCl}
T_n(k) & = & \left\{ ({\bf x}_1, {\bf x}_2) \in {\cal X}_1^n \times {\cal X}_2^n \left| - \log P_{X_1^{(k)n}|X_2^{(k)n}}({\bf x}_1|{\bf x}_2) < \sqrt{n}\left( z_n^{(1)} -2\gamma_n \right),  \right. \right. \nonumber \\
& & \hspace*{3.2cm} - \log P_{X_2^{(k)n}|X_1^{(k)n}}({\bf x}_2|{\bf x}_1) < \sqrt{n}\left( z_n^{(2)} -2\gamma_n \right), \nonumber \\
& & \hspace*{3.2cm} \left. - \log P_{X_1^{(k)n}X_2^{(k)n}}({\bf x}_1,{\bf x}_2) < \sqrt{n}\left( z_n^{(3)} -2\gamma_n \right) \right\}.
\end{IEEEeqnarray*}

Summarizing (\ref{delta1})--(\ref{G3}) yields
\begin{IEEEeqnarray}{rCl} \label{eq:appc-1}
P_{X_1^{(k)n}X_2^{(k)n}}(G_n(k)|E_n) & \geq & P_{X_1^{(k)n}X_2^{(k)n}}(T_n(k)|E_n).
\end{IEEEeqnarray}

On the other hand, by means of (\ref{eq:appc-0}) and (\ref{eq:appc-1}),
\begin{IEEEeqnarray}{rCl}  \label{eq:appc-2}
P_{X_1^{(k)n}X_2^{(k)n}}(S_n) & \geq & P_{X_1^{(k)n}X_2^{(k)n}}(T_n(k)|E_n)P_{X_1^{(k)n}X_2^{(k)n}}(E_n) \nonumber \\
& = & P_{X_1^{(k)n}X_2^{(k)n}}(T_n(k)) - P_{X_1^{(k)n}X_2^{(k)n}}(T_n(k)|E_n^c)P_{X_1^{(k)n}X_2^{(k)n}}(E_n^c) \nonumber \\
& \geq & P_{X_1^{(k)n}X_2^{(k)n}}(T_n(k)) - 2e^{-\sqrt{n}\gamma_n},
\end{IEEEeqnarray}
for sufficiently large $n$ where $\lq\lq c"$ denotes the complement, and in the last step we have used (\ref{eq:appc-3}).
Finally, it suffices only to replace $2\gamma_n$ by $\gamma_n$ and to notice that (\ref{eq:appc-2}) gives the second inequality of the lemma.
\IEEEQED
\section*{Acknowledgment}
The authors are very grateful to Hideki Yagi for useful discussions which led to improve Lemma \ref{exponent}.
\ifCLASSOPTIONcaptionsoff
  \newpage
\fi


\begin{thebibliography}{10}
\providecommand{\url}[1]{#1}
\csname url@samestyle\endcsname
\providecommand{\newblock}{\relax}
\providecommand{\bibinfo}[2]{#2}
\providecommand{\BIBentrySTDinterwordspacing}{\spaceskip=0pt\relax}
\providecommand{\BIBentryALTinterwordstretchfactor}{4}
\providecommand{\BIBentryALTinterwordspacing}{\spaceskip=\fontdimen2\font plus
\BIBentryALTinterwordstretchfactor\fontdimen3\font minus
  \fontdimen4\font\relax}
\providecommand{\BIBforeignlanguage}[2]{{%
\expandafter\ifx\csname l@#1\endcsname\relax
\typeout{** WARNING: IEEEtran.bst: No hyphenation pattern has been}%
\typeout{** loaded for the language `#1'. Using the pattern for}%
\typeout{** the default language instead.}%
\else
\language=\csname l@#1\endcsname
\fi
#2}}
\providecommand{\BIBdecl}{\relax}
\BIBdecl

\bibitem{HV93}
T.~S. Han and S.~Verd\'u, ``Approximation theory of output statistics,''
  \emph{{IEEE} Trans. Inf. Theory}, vol.~39, no.~3, pp. 752--772, 1993.

\bibitem{Steinberg}
Y.~Steinberg and S.~Verd\'u, ``Simulation of random processes and
  rate-distortion theory,'' \emph{{IEEE} Trans. Inf. Theory}, vol.~42, no.~1,
  pp. 63--86, 1996.

\bibitem{Kon}
I.~Kontoyiannis, ``Second-order noiseless source coding theorems,''
  \emph{{IEEE} Trans. Inf. Theory}, vol.~43, no.~4, pp. 1339--1341, 1997.

\bibitem{Strassen}
V.~Strassen, ``Asymptotische absh{\"a}tzungen in {S}hannon's informations
  theorie,'' in \emph{Trans. 3rd Prague Conf. Inf. Theory}, 1962, pp. 687--723.

\bibitem{CK}
I.~Csisz\'{a}r and J.~{K\"{o}rner}, \emph{Information Theory: Coding Theorems
  for Discrete Memoryless Systems}.\hskip 1em plus 0.5em minus 0.4em\relax
  Academic, 1981.

\bibitem{Hayashi2}
M.~Hayashi, ``Information spectrum approach to second-order coding rate in
  channel coding,'' \emph{{IEEE} Trans. Inf. Theory}, vol.~55, no.~11, pp.
  4947--4966, 2009.

\bibitem{Polyanskiy2010}
Y.~Polyanskiy, H.~Poor, and S.~Verd\'u, ``Channel coding rate in the finite
  blocklength regime,'' \emph{{IEEE} Trans. Inf. Theory}, vol.~56, no.~5, pp.
  2307--2359, 2010.

\bibitem{Hayashi}
M.~Hayashi, ``Second-order asymptotics in fixed-length source coding and
  intrinsic randomness,'' \emph{{IEEE} Trans. Inf. Theory}, vol.~54, no.~10,
  pp. 4619--4637, 2008.

\bibitem{Han}
T.~S. Han, \emph{Information-Spectrum Methods in Information Theory}.\hskip 1em
  plus 0.5em minus 0.4em\relax Springer, New York, 2003.

\bibitem{NH2011}
R.~Nomura and T.~S. Han, ``Second-order resolvability, intrinsic randomness,
  and fixed-length source coding for mixed sources: information spectrum
  approach,'' \emph{{IEEE} Trans. Inf. Theory}, vol.~59, no.~1, pp. 1--16,
  2013.

\bibitem{Cover}
T.~M. Cover and J.~A. Thomas, \emph{Elements of Information Theory}.\hskip 1em
  plus 0.5em minus 0.4em\relax Wiley, 1991.

\bibitem{SW}
D.Slepian and J.K.Wolf, ``Noiseless coding of correlated information sources,''
  \emph{{IEEE} Trans. Inf. Theory}, vol.~19, no.~4, pp. 471--480, 1973.

\bibitem{Steinberg94}
Y.~Steinberg and S.~Verd\'u, ``Channel simulation and coding with side
  information,'' \emph{{IEEE} Trans. Inf. Theory}, vol.~40, no.~3, pp.
  634--646, 1994.

\bibitem{WMU2010}
S.~Watanabe, R.~Matsumoto, and T.~Uyematsu, ``Strongly secure privacy
  amplification cannot be obtained by encoder of {Slepian}-{W}olf code,''
  \emph{IEICE Trans. Fundamentals}, vol. E93-A, no.~9, pp. 1650--1659, 2010.

\bibitem{Nomura_SITA2009}
R.~Nomura and T.~Matsushima, ``A note on the second order separate source
  coding theorem for sources with side information,'' in \emph{Proc. The 32nd
  Symposium on Information Theory and its Applications}, 2009, pp. 637--642.

\bibitem{MK}
S.~Miyake and F.~Kanaya, ``Coding theorems on correlated general sources,''
  \emph{IEICE Trans. Fundamentals}, vol. E78-A, no.~9, pp. 1063--1070, 1995.

\bibitem{Nomura2011_2}
R.~Nomura and T.~Matsushima, ``An analysis of {Slepian}-{Wolf} coding problem
  based on the asymptotic normality,'' \emph{IEICE Trans. Fundamentals}, vol.
  E94-A, no.~11, pp. 2220--2225, 2011.

\bibitem{Tan_ISIT2012}
V.~Y.~F. Tan and O.~Kosut, ``The dispersion of {Slepian}-{Wolf} coding,'' in
  \emph{Proc. 2012 IEEE International Symposium on Information Theory}, 2012,
  pp. 920--924.

\bibitem{Feller}
W.~Feller, \emph{An Introduction to Probability Theory and Its Applications,
  Vol.1 and 2}.\hskip 1em plus 0.5em minus 0.4em\relax John Willey and Sons,
  New York, 1966.

\bibitem{Sazonov}
V.~V. Sazonov, \emph{Normal Approximations: Some Recent Advances}.\hskip 1em
  plus 0.5em minus 0.4em\relax Springer, New York, 1981.

\bibitem{Bentkus}
V.~Bentkus, ``On the dependence of the {B}erry-{E}sseen bound on dimension,''
  \emph{J. Stat. Planning and Inference}, vol. 113, no.~2, pp. 385--402, 2003.

\bibitem{Billingsley}
P.~Billingsley, \emph{Probability and Measure}.\hskip 1em plus 0.5em minus
  0.4em\relax Wiley, 1995.

\bibitem{Koshelev}
V.~N. Koshelev, ``On a problem of separate coding of two dependent sources,''
  \emph{Problems of Information Transmission}, vol.~13, no.~1, pp. 18--22,
  1977.

\bibitem{Gallager}
R.~G. Gallager, \emph{Information Theory and Reliable Communication}.\hskip 1em
  plus 0.5em minus 0.4em\relax Wiley, 1968.

\bibitem{Polyanskiy2011}
Y.~Polyanskiy, H.~V. Poor, and S.~Verd\'{u}, ``{Dispersion of the
  Gilbert-Elliott Channel},'' \emph{{IEEE} Trans. Inf. Theory}, vol.~57, no.~4,
  pp. 1829--1848, 2011.

\end{thebibliography}
\end{document}